\documentclass[lettersize,journal]{IEEEtran}
\usepackage{amsmath,amsfonts}
\usepackage{algorithmic}
\usepackage{algorithm}
\usepackage{array}
\usepackage{textcomp}
\usepackage{stfloats}
\usepackage{url}
\usepackage{verbatim}
\usepackage{graphicx}
\usepackage{cite}

\usepackage{tabularx}
\usepackage{makecell}
\usepackage{caption}
\usepackage{subcaption}
\usepackage{booktabs}
\usepackage{lipsum}
\usepackage[switch]{lineno}

\begin{document}

\title{NDT: Non-Differential Transformer and Its Application to Sentiment Analysis}

\author{Soudeep Ghoshal,
Himanshu Buckchash,
Sarita Paudel,
Rub\'{e}n Ruiz-Torrubiano
\thanks{SG is with Kalinga Institute of Industrial Technology (KIIT), Bhubaneswar, Odisha 751024, India. HB, SP, RRT are with IMC University of Applied Sciences Krems, Piaristengasse 1, 3500 Austria.}%
\thanks{}}

\markboth{}%
{Ghoshal \MakeLowercase{\textit{et al.}}: NDT: Non-Differential Transformer and Its Application to Sentiment Analysis}

\maketitle

\begin{abstract}
From customer feedback to social media, understanding human sentiment in text is central to how machines can interact meaningfully with people.
However, despite notable progress, accurately capturing sentiment remains a challenging task, which continues to motivate further research in this area. To this end, we introduce Non-Differential Transformer (NDT). It is inspired by (but in contrast to) the state-of-the-art Differential Transformer (DT) model. While standard Transformers can struggle with irrelevant context, the sota DT model uses attention map subtraction, potentially for noise cancellation. We explore an alternative motivation, hypothesizing that benefits may arise from enabling different attention components to specialize on distinct concepts within the text, similar to multiplexing information channels or mixture models, rather than primarily canceling noise via subtraction. Guided by this concept-multiplexing (ConPlex) view, the specific architecture presented in this paper employs a purely additive strategy. It uses only positive weights, learned during training, to ensure constructive combination of these specialized attention perspectives. This design choice explores positive only integration, though our broader framework also shows promise with less constrained linear combinations involving both positive and negative weights. Our model computes attention via this positively weighted sum of multiple distinct attention maps. This allows the model to constructively integrate diverse signals and potentially capture more complex contextual relationships. Competitive performance is achieved by the proposed model for Sentiment Analysis while tested on multiple datasets. We conclude by presenting our results, challenges and future research agenda in this important area of research.
\end{abstract}


\section{Introduction}
\IEEEPARstart{U}{nderstanding} human sentiment in text is central to how machines interact meaningfully with people. Sentiment analysis has become critical for extracting insights from vast amounts of unstructured data~\cite{liu2022sentiment, giachanou2016like}, from e-commerce platforms processing millions of product reviews to financial institutions analyzing market sentiment for trading decisions~\cite{loughran2011liability, bollen2011twitter}. Despite significant advances in natural language processing, accurately capturing sentiment remains challenging. Traditional approaches, from lexicon based methods to classical machine learning with handcrafted features, have demonstrated fundamental limitations in handling the nuanced complexities of human emotional expression~\cite{taboada2011lexicon}. While deep learning and transformer architectures like BERT~\cite{devlin2019bert} have substantially advanced the field, these improvements have focused primarily on scaling model capacity rather than addressing fundamental architectural limitations in how attention mechanisms process sentiment bearing information.

The transformer architecture's multi head attention mechanism faces inherent challenges in sentiment analysis tasks~\cite{vaswani2017attention}. Standard transformers struggle with irrelevant context, particularly in longer texts where sentiment bearing elements may be dispersed among substantial neutral or contradictory content. The Differential Transformer (DT)~\cite{ye2025differential} addresses this through attention map subtraction operating under noise cancellation principles:
\[
Attention = softmax(A_0) - \lambda \cdot softmax(A_1)
\]
where $\lambda$ is a learned positive coefficient~\cite{ye2025differential}, and $A_0, A_1$ are the first and second attention score matrices, computed from first and second pair of projected Q and K~\cite{ye2025differential}. While DT has demonstrated improvements, we propose an alternative theoretical framework challenging this noise centric view. We hypothesize that benefits arise not from noise cancellation, but from enabling different attention components to specialize on distinct conceptual aspects within text, drawing inspiration from multiplexing techniques in signal processing and mixture of experts frameworks in machine learning~\cite{jacobs1991adaptive, shazeer2017outrageously}.

We introduce concept-multiplexing (ConPlex), which posits that effective sentiment understanding requires simultaneous attention to multiple conceptual channels: lexical sentiment indicators, contextual modifiers, syntactic structures, and domain specific patterns. Rather than subtracting potentially useful information, constructive combination of specialized attention perspectives can yield superior contextual understanding. Building upon ConPlex, we present the Non-Differential Transformer (NDT), employing purely additive attention combinations. The NDT generalizes the binary differential formula to support $N$ attention components through the formulation:
\[
\begin{split}
Attention = & \; softmax(A_0) + \lambda_1 \cdot softmax(A_1) \\
& + \dots + \lambda_{N-1} \cdot softmax(A_{N-1}),
\end{split}
\]
where each $\lambda_i$ represents a learnable coefficient controlling the $i^{\text{th}}$ specialized attention component's contribution. This design explores positive only integration, though our broader framework demonstrates effectiveness with various coefficient constraints including both positive and negative weights.

Our approach addresses key limitations: (1) information preservation through additive combinations, (2) explicit component specialization through distinct parameterizations, (3) flexible integration of multiple sentiment indicators through learned weighting schemes, and (4) computational efficiency through shared value projections.

\textbf{Contributions:} \textbf{(a)} First, we introduce the ConPlex theoretical framework providing an alternative foundation for multicomponent attention mechanisms, contradicting the noise cancellation hypothesis~\cite{ye2025differential}. \textbf{(b)} Second, we present the NDT architecture operationalizing ConPlex through additive attention with learnable coefficients. \textbf{(c)} Third, we conduct comprehensive experiments across multiple sentiment datasets (IMDB, SST-2, YELP Reviews, Twitter Financial News) with systematic evaluation of coefficient constraints ($[0,1]$, $[-1,1]$, $[0,\infty)$, and $(-\infty,\infty)$) and component scaling.
\textbf{(d)} Fourth, we analyze component specialization through lambda evolution and attention visualization.

Our results demonstrate consistent improvements over standard transformers and DT: up to 1.36\% on IMDB, 0.32\% on SST-2, 0.32\% on YELP Reviews, and 1.00\% on Twitter Financial News. Analysis of learned lambda values provides evidence for component specialization, with different constraints leading to distinct attention patterns suitable for various sentiment understanding requirements. Section~\ref{section:related_works} reviews related work, Section~\ref{section:method} describes methodology, Section~\ref{section:results} presents comprehensive results and analysis and finally, Section~\ref{section:conclusion} presents the conclusion.

\section{Related Works} \label{section:related_works}

\subsection{Evolution of Sentiment Analysis Approaches}

Sentiment analysis has evolved from lexicon based methods through classical machine learning to modern deep learning architectures. Early approaches utilized sentiment dictionaries to assign polarity scores but struggled with context dependent expressions~\cite{taboada2011lexicon}. Classical methods including SVM and Naive Bayes employed handcrafted features yet demonstrated limitations in capturing nuanced emotional expression.

The advent of deep learning introduced CNNs and LSTMs for hierarchical representation learning. Hybrid CNN-LSTM architectures emerged as particularly effective, with CNNs extracting local features and LSTMs capturing long range dependencies~\cite{tan2022roberta, lai2015recurrent}. Transformer based models like BERT~\cite{devlin2019bert} revolutionized the field through pretrained contextual embeddings, with subsequent variants like RoBERTa~\cite{liu2019roberta} and XLNet~\cite{yang2019xlnet} further improving performance through enhanced training procedures. Despite these advances, persistent challenges remain in domain adaptation, and multi class classification with fine grained rating scales, motivating continued architectural innovations beyond simple capacity scaling.

\subsection{Attention Mechanisms in Transformer Architectures}

The transformer architecture introduced by Vaswani \textit{et al.}~\cite{vaswani2017attention} replaced recurrence with pure attention mechanisms, computing attention through query-key interactions scaled by key dimension. Multi head attention extends this by learning multiple parallel attention functions with different parameterizations, enabling joint attention to information from different representation subspaces. The parallelizability of these operations enables efficient training compared to sequential recurrent architectures.

Subsequent research explored numerous variants including sparse attention methods that reduce computational complexity for long sequences. Longformer~\cite{beltagy2020longformer} introduced sliding window attention combined with global tokens, while BigBird~\cite{zaheer2020big} added random attention to maintain expressivity while achieving linear complexity. Studies on attention interpretability revealed that different heads naturally specialize on distinct patterns (syntactic relationships, positional information, or semantic connections)~\cite{voita2019analyzing}, providing empirical motivation for mechanisms that explicitly promote component specialization, as explored in the NDT architecture.

\subsection{Differential Transformer and Subtractive Attention Mechanisms}

The Differential Transformer by Ye \textit{et al.}~\cite{ye2025differential} represents a significant departure from standard attention through its subtractive approach:
$\text{Attention}_{diff} = softmax(A_0) - \lambda \cdot softmax(A_1)$, where $\lambda>0$
controls subtraction magnitude. This design stems from a noise cancellation hypothesis proposing that subtracting two attention maps filters common mode noise while amplifying relevant signals, drawing inspiration from differential amplifiers in electrical engineering.

The underlying philosophy posits that two attention components capture similar irrelevant context (noise) but different relevant information (signal), such that subtraction preferentially removes noise while preserving signal through destructive interference. While demonstrating empirical success in language modeling and long context understanding, this noise centric view merits examination. The NDT presents an alternative framework challenging this perspective, hypothesizing that benefits arise from enabling attention components to specialize on distinct conceptual aspects with constructive rather than destructive combination. This fundamental philosophical difference (viewing components as complementary information channels requiring integration rather than opposing signals requiring cancellation) distinguishes ConPlex from subtractive approaches.

\subsection{Component Specialization and Mixture of Experts in Neural Networks}

The principle of component specialization has rich foundations in mixture of experts (MoE) frameworks. Jacobs \textit{et al.}~\cite{jacobs1991adaptive} demonstrated that training multiple expert networks on different input subsets, coordinated by a gating network, enables more efficient learning through natural specialization. Modern implementations by Shazeer \textit{et al.}~\cite{shazeer2017outrageously} scale to billions of parameters through sparse gating, where each input routes to small expert subsets, achieving dramatic capacity increases without proportional computational growth.

These specialization principles extend to attention mechanisms. Rather than treating attention as monolithic, decomposing it into specialized components, each potentially focusing on different conceptual channels like lexical indicators, syntactic structures, or contextual modifiers, could yield more powerful representations. Multi head attention exhibits natural specialization, though standard implementations use simple concatenation rather than learned weighted integration. The ConPlex framework draws inspiration from these principles, hypothesizing that explicit parameterization of multiple attention components with learned combination weights promotes stronger specialization. Unlike MoE's discrete routing, NDT employs continuous learned coefficients enabling all components to contribute while maintaining specialization through distinct parameterizations, providing a middle ground between sparse routing and uniform weighting.

\section{Method} \label{section:method}

This section presents the Non-Differential Transformer (NDT) architecture, beginning with the theoretical foundation of concept-multiplexing (ConPlex) and proceeding to detailed architectural specifications. We systematically develop the mathematical formulations, implementation strategies, and training methodologies that enable effective sentiment analysis through additive attention mechanisms.

\subsection{Concept-Multiplexing (ConPlex) Framework}

The ConPlex framework represents a fundamental departure from noise cancellation approaches to attention mechanisms. We hypothesize that effective sentiment understanding emerges from simultaneous processing of multiple conceptual channels within text, each requiring specialized attention patterns for optimal comprehension.

\subsubsection{Theoretical Foundation}

Let $S(x)$ represent the sentiment understanding function for input sequence $x$. Traditional attention mechanisms model this as a single unified attention pattern:
\(
S(x) = \text{softmax}\!\left(\frac{QK^{T}}{\sqrt{d}}\right) \cdot V
\).
The Differential Transformer~\cite{ye2025differential} extends this through subtractive combination:

\(
S_{\text{diff}}(x) = 
\left[
\text{softmax}\!\left(\frac{Q_{0}K_{0}^{T}}{\sqrt{d}}\right)
- \lambda \cdot \text{softmax}\!\left(\frac{Q_{1}K_{1}^{T}}{\sqrt{d}}\right)
\right] \cdot V
\), 
where $\lambda > 0$ enforces noise cancellation through subtraction.

The ConPlex framework generalizes this to $N$ specialized components through additive combination:
\begin{equation}
S_{\text{ConPlex}}(x) =
\left[
\sum_{i=0}^{N-1} \lambda_i \cdot 
\text{softmax}\!\left(\frac{Q_iK_i^{T}}{\sqrt{d}}\right)
\right] \cdot V
\end{equation}

where $\lambda_0$ = 1 is fixed and $\lambda_i$ are learnable coefficients that enable constructive integration of specialized attention perspectives. We hypothesize that each attention component $A_i$ specializes on distinct conceptual channels relevant to sentiment analysis: lexical sentiment indicators (e.g., ``excellent'' ``terrible''), contextual modifiers (e.g., ``not bad,'' ``surprisingly good'')~\cite{poria2014sentic}, syntactic structures that convey sentiment, and domain specific patterns, with $\lambda_i$ coefficients reflecting the relative importance of each specialization.

\subsection{Non-Differential Transformer Architecture}

\subsubsection{Multi Component Attention Mechanism}

The core innovation of NDT lies in its generalized attention computation that supports arbitrary numbers of specialized components. For $N$ attention components, each component $i$ maintains separate query and key projections while sharing a common value projection:

\begin{equation}
Q_i = XW_{Q_i}, \quad 
K_i = XW_{K_i}, \quad 
V = XW_V
\end{equation}

where $W_{Q_i} \in \mathbb{R}^{d \times d/2}$, 
$W_{K_i} \in \mathbb{R}^{d \times d/2}$, and 
$W_V \in \mathbb{R}^{d \times d}$.

The reduced dimensionality $(d/2)$ for individual $Q$ and $K$ projections 
encourages component specialization while maintaining computational efficiency. 
For multi head attention with $H$ heads, each component's head dimension becomes 
$d_{\text{head}} = (d/2)/{H}$. The attention computation for each head $h$ becomes:
\begin{equation}
\text{Attention}_{h} = 
\sum_{i=0}^{N-1} \lambda_i \cdot 
\text{softmax}\!\left(\frac{Q_{i,h} K_{i,h}^T}{\sqrt{d_{\text{head}}}}\right) \cdot V_h
\end{equation}

\subsubsection{Lambda Parameter Learning}
The $\lambda_i$ coefficients are learned through interaction based computation 
that combines query and key parameters. Each component $i$ maintains learnable 
parameters $\lambda_{Q_i} \in \mathbb{R}^{d/4}$ and 
$\lambda_{K_i} \in \mathbb{R}^{d/4}$, where the lambda computation proceeds as:
\begin{equation}
\lambda'_{Q_i} = f_{\text{constraint}}(\lambda_{Q_i}), \quad 
\lambda'_{K_i} = f_{\text{constraint}}(\lambda_{K_i}),
\end{equation}
\begin{equation}
\text{interaction}_i = \text{mean}\!\left(\lambda'_{Q_i} \odot \lambda'_{K_i}\right),
\end{equation}
\begin{equation}
\lambda_i = \text{interaction}_i \cdot \alpha_{\text{init},i} 
+ (1 - \text{interaction}_i) \cdot \beta_i,
\end{equation}

where $\odot$ denotes element wise multiplication, 
$\alpha_{\text{init},i}$ is a component specific initialization value, 
and $\beta_i$ is a component specific bias term. 

\subsubsection{Constraint Configurations}

We investigate four constraint configurations for $\lambda_i$ parameters:

\begin{itemize}
    \item \textbf{Bounded Positive $[0,1]$:}
    \(f_{\text{constraint}}(x) = \sigma(x)\)
    where $\sigma$ is the sigmoid function, ensuring purely constructive combinations.

    \item \textbf{Symmetric $[-1,1]$:}
    \(f_{\text{constraint}}(x) = \tanh(x)\)
    allowing both positive and negative contributions for selective emphasis/suppression.

    \item \textbf{Non-negative $[0,\infty)$:}
    \(f_{\text{constraint}}(x) = \text{ReLU}(x)\)
    providing unbounded positive combinations.

    \item \textbf{Unconstrained $(-\infty,\infty)$:}
    \(f_{\text{constraint}}(x) = x\)
    offering maximum learning flexibility.
\end{itemize}

\subsection{Architectural Integration}

\subsubsection{Multi Head Extension}

The NDT mechanism extends naturally to multi head attention. For $H$ attention heads, each component $i$ is reshaped to support multi head computation with head dimension $d_{\text{head}} = (d/2)/H$:
\begin{equation}
\text{head}_h = 
\sum_{i=0}^{N-1} \lambda_i \cdot 
\text{softmax}\!\left(\frac{Q_{i,h} K_{i,h}^T}{\sqrt{d_{\text{head}}}}\right) \cdot V_h
\end{equation}

The final multi head output concatenates all heads:
\[
\text{MultiHead}(Q,K,V) = \text{Concat}(\text{head}_1, \dots, \text{head}_H) W^O,
\]
where each component maintains the same lambda value $\lambda_i$ across all heads within a given layer.

\subsubsection{Layer Integration}

Each NDT layer follows the standard transformer encoder structure with pre layer normalization:
\begin{equation}
Y = X + \text{NDT-Attention}(\text{RMSNorm}(X)),
\end{equation}
\begin{equation}
Z = Y + \text{FFN}(\text{RMSNorm}(Y)),
\end{equation}

We employ RMSNorm instead of LayerNorm for improved training stability, and the feed forward network uses SwiGLU activation for enhanced expressivity.

\subsubsection{Component Scaling Strategy}
We systematically evaluate component counts of $N \in \{2, 3, 4\}$ to analyze the relationship between architectural complexity and performance. For the base component variant, the attention computation becomes:
\begin{equation}
\begin{split}
\text{Attention} = & \; \text{softmax}\!\left(\frac{Q_0 K_0^T}{\sqrt{d/2}}\right) \cdot V \\
& + \sum_{j=1}^{N-1} \lambda_j \cdot 
\text{softmax}\!\left(\frac{Q_j K_j^T}{\sqrt{d/2}}\right) \cdot V
\end{split}
\end{equation}

This formulation maintains a fixed base component while allowing additional components to provide specialized contributions.

\subsection{Training Methodology}

\subsubsection{Parameter Optimization}

NDT employs a multi group optimization strategy using AdamW with different weight decay rates: lambda parameters use \texttt{weight\_decay = 0.0} (no regularization) to learn freely, while architectural parameters use \texttt{weight\_decay = 0.1} (standard L2 penalty). The training objective uses standard crossentropy loss: \(L = \text{CrossEntropy}(\hat{y}, y)\).

\subsubsection{Initialization Strategy}

Lambda parameters are initialized to promote component diversity while ensuring training stability. The raw lambda parameters $\lambda_{Q_i}$ and $\lambda_{K_i}$ are initialized from a normal distribution $\mathcal{N}(0, 0.02^2)$, with component specific initialization targets $\alpha_{\text{init},i}$ computed based on layer depth and component index. These values are constraint specific and clamped to appropriate ranges for each variant. Architectural parameters follow standard transformer initialization with Xavier uniform distribution for weight matrices and zero initialization for biases.

\subsection{Evaluation Metrics}

To comprehensively assess NDT performance, we employ multiple complementary evaluation metrics spanning classification accuracy, representation quality, and computational efficiency.

\textbf{Classification Performance:} We use standard metrics including accuracy (proportion of correct predictions), precision ($\text{TP} / (\text{TP} + \text{FP})$), F1 score (harmonic mean of precision and recall), and Area Under the ROC Curve (AUC) for discriminative performance across classification thresholds.

\textbf{Representation Quality:} We quantify learned embedding quality using the silhouette score, which measures how well samples cluster with their true class relative to other classes. For each sample $i$, the silhouette coefficient is:
\[
s(i) = \frac{b(i) - a(i)}{\max(a(i), b(i))}
\]
where $a(i)$ is the average intracluster distance and $b(i)$ is the minimum average intercluster distance. Values range from $-1$ to $1$, with higher values indicating better defined clusters with greater separation between classes.

\textbf{Computational Efficiency:} We report total trainable parameters (in millions) as well as the difference relative to the Vanilla Transformer baseline:
\(\mbox{
$\Delta \text{Params (M)} = \text{Params}_{\text{NDT}} - \text{Params}_{\text{Vanilla}}$
}\),
along with average inference time (in milliseconds) per test set batch for each dataset, measured on identical hardware configurations.

\section{Results and Discussions} \label{section:results}

This section presents comprehensive experimental results evaluating the Non-Differential Transformer across four sentiment analysis datasets. We systematically compare NDT variants against baseline transformers, analyzing performance metrics, computational efficiency, and learned representations.

\subsection{Experimental Setup}
\subsubsection{Datasets}
We evaluate on four sentiment analysis benchmarks with varying characteristics:

\noindent
\textbf{IMDB Movie Reviews:} Binary sentiment classification with 20,000 training, 5,000 validation, and 25,000 test samples. Vocabulary size: 26,284 tokens \cite{maas2011learning}.\\
\textbf{SST-2:} Binary sentiment classification with 47,754 training, 10,233 validation, and 10,234 test samples. Vocabulary size: 8,865 tokens \cite{socher2013recursive}.\\
\textbf{YELP Reviews:} 5 class rating prediction with 489,999 training, 105,001 validation, and 105,001 test samples. Vocabulary size: 63,385 tokens \cite{yelpdataset}.\\
\textbf{Twitter Financial News:} Ternary sentiment classification with 9,543 training, 1,194 validation, and 1,194 test samples. Vocabulary size: 3,137 tokens \cite{zeroshottwit}.

\subsubsection{Model Configurations}
For each dataset, we evaluate 14 model configurations:

\noindent
\textbf{Baselines:} Vanilla Transformer~\cite{vaswani2017attention},  Differential Transformer~\cite{ye2025differential}.\\
\textbf{NDT Variants:} Four constraint configurations ($[0,1]$, $[-1,1]$, $[0,\infty)$, $(-\infty,\infty)$) $\times$ Three component counts (2, 3, 4).

\subsection{IMDB Movie Reviews Results}

\subsubsection{Overall Performance Comparison}

The NDT framework demonstrates consistent performance improvements over both baseline architectures on the \textit{IMDB} dataset (Table~\ref{tab:acc_imdb}). The \textbf{NDT [0,1]} variant with 2 components achieves the highest test accuracy of \textbf{85.56\%}, representing improvements of \textbf{+1.57\%} over the Vanilla Transformer and \textbf{+1.36\%} over the Differential Transformer. This purely additive, positive weight configuration outperforms the subtractive DT approach, providing empirical support for the \textit{ConPlex hypothesis} that constructive integration of specialized attention components can be more effective than noise cancellation through subtraction.

Across all constraint configurations, the 2 component variants generally achieve the strongest performance, with the \textbf{[0,1]}, \textbf{[-1,1]}, \textbf{[0,$\infty$)}, and \textbf{($-\infty$,$\infty$)} configurations yielding accuracies of 85.56\%, 85.33\%, 85.13\%, and 85.19\%, respectively. The diminishing returns observed with 3 and 4 components (with the notable exception of NDT [0,1] with 4 components at 85.50\%) suggest that two specialized attention perspectives may be sufficient for capturing the primary conceptual channels in binary sentiment classification tasks.

Notably, 11 out of 12 NDT variants surpass the DT baseline, with precision and F1 score metrics showing similar improvement patterns, indicating that the performance gains extend beyond simple accuracy to overall classification quality. AUC scores remain competitive across NDT variants, though DT retains a marginal advantage in ranking ability on this binary task.

\begin{table}[h!]
\centering
\caption{Accuracy and Classification Metrics on IMDB Dataset}
\label{tab:acc_imdb}
\begin{tabular}{lcccc}
\hline
$N_{comp}$ & Acc (\%) & Precision & F1 Score & AUC \\
\hline
\multicolumn{5}{c}{\textbf{Vanilla Transformer}} \\
-- & 83.99 & 0.844 & 0.839 & 0.927 \\
\hline
\multicolumn{5}{c}{\textbf{Differential Transformer}} \\
2 & 84.20 & 0.851 & 0.841 & 0.935 \\
\hline
\multicolumn{5}{c}{\textbf{NDT [0,1]}} \\
2 & 85.56 & 0.856 & 0.856 & 0.933 \\
3 & 84.74 & 0.851 & 0.847 & 0.931 \\
4 & 85.50 & 0.855 & 0.855 & 0.932 \\
\hline
\multicolumn{5}{c}{\textbf{NDT [-1,1]}} \\
2 & 85.33 & 0.854 & 0.853 & 0.931 \\
3 & 85.28 & 0.853 & 0.853 & 0.932 \\
4 & 83.69 & 0.837 & 0.837 & 0.917 \\
\hline
\multicolumn{5}{c}{\textbf{NDT [0,$\infty$)}} \\
2 & 85.13 & 0.852 & 0.851 & 0.932 \\
3 & 84.75 & 0.851 & 0.847 & 0.932 \\
4 & 84.68 & 0.847 & 0.847 & 0.927 \\
\hline
\multicolumn{5}{c}{\textbf{NDT ($-\infty$,$\infty$)}} \\
2 & 85.19 & 0.852 & 0.852 & 0.931 \\
3 & 84.68 & 0.852 & 0.848 & 0.932 \\
4 & 84.73 & 0.847 & 0.847 & 0.926 \\
\hline
\end{tabular}
\end{table}

\subsubsection{Lambda Value Analysis}

The learned $\lambda$ values reveal distinct patterns across constraint configurations (Table~\ref{tab:lambda_imdb}). The \textbf{NDT [0,1]} variants show consistently positive, balanced contributions (0.1365 - 0.1860) with decreasing trends across component indices, suggesting all components provide constructive information. The \textbf{NDT [-1,1]} configuration exhibits exclusively negative $\lambda$ values (-0.2001 to -0.0575), indicating learned subtractive combinations even when positive weights are available, though this performs below the purely positive [0,1] configuration. The unbounded variants \textbf{[0,$\infty$)} and \textbf{($-\infty$,$\infty$)} demonstrate higher magnitudes (up to 0.3786 in Layer~2), with $\lambda$ values consistently increasing from Layer~1 to Layer~2, suggesting deeper layers benefit from stronger integration. The hierarchical progression across components (e.g., 0.2192~$\rightarrow$~0.1803~$\rightarrow$~0.1461) demonstrates learned importance ordering, supporting the hypothesis that the model appropriately weights different conceptual channels during training.

\begin{table}[h!]
\centering
\caption{Learned Lambda Values for the IMDB Dataset}
\label{tab:lambda_imdb}
\begin{tabular}{lcc}
\hline
$N_{comp}$ & \makecell{Layer 1\\$\lambda$ Values}          & \makecell{Layer 2\\$\lambda$ Values} \\
\hline
\multicolumn{3}{c}{\textbf{NDT [0,1]}} \\
2 & (0.1507)                 & (0.1860) \\
3 & (0.1506, 0.1436)         & (0.1841, 0.1604) \\
4 & (0.1515, 0.1436, 0.1365) & (0.1845, 0.1603, 0.1513) \\
\hline
\multicolumn{3}{c}{\textbf{NDT [-1,1]}} \\
2 & (-0.2001)                   & (-0.0575) \\
3 & (-0.1997, -0.1599)          & (-0.0650, -0.0958) \\
4 & (-0.2000, -0.1604, -0.1282) & (-0.0596, -0.0921, -0.0708) \\
\hline
\multicolumn{3}{c}{\textbf{NDT [0,$\infty$)}} \\
2 & (0.2178)                 & (0.3756) \\
3 & (0.2191, 0.1781)         & (0.3713, 0.2193) \\
4 & (0.2192, 0.1803, 0.1461) & (0.3677, 0.2222, 0.1793) \\
\hline
\multicolumn{3}{c}{\textbf{NDT ($-\infty$,$\infty$)}} \\
2 & (0.2083)                 & (0.3786) \\
3 & (0.2146, 0.1760)         & (0.3575, 0.2087) \\
4 & (0.2097, 0.1810, 0.1406) & (0.3556, 0.2155, 0.1711) \\
\hline
\end{tabular}
\end{table}

\subsection{SST-2 Results}

\subsubsection{Overall Performance Comparison}
Similar to IMDB, the NDT framework demonstrates consistent improvements over baseline architectures on the \textit{SST-2} dataset (Table~\ref{tab:acc_sst-2}). The \textbf{NDT [0,1]} variant with 3 components achieves the highest test accuracy of \textbf{91.85\%}, representing improvements of \textbf{+1.00\%} over the Vanilla Transformer and \textbf{+0.32\%} over the Differential Transformer. 

Unlike IMDB, where 2 components proved optimal, SST-2 shows the best performance with 3 components in the [0,1] configuration, suggesting that the shorter, more syntactically complex sentences in SST-2 may benefit from an additional specialized attention perspective. The \textbf{NDT [-1,1]} variant with 2 components also performs strongly at 91.83\%, closely matching the best result. 

Across all constraint configurations, performance remains tightly clustered between 91.06\% and 91.85\%, indicating more stable behavior across variants compared to IMDB’s wider performance spread. The superior AUC score of \textbf{0.985} achieved by \textbf{NDT [0,$\infty$)} with 4 components, despite lower accuracy (91.06\%), suggests that this configuration excels at ranking predictions even when classification thresholds are suboptimal. 

Overall, 6 out of 12 NDT variants surpass the DT baseline, reinforcing the effectiveness of the \textit{ConPlex framework} across different dataset characteristics.

\begin{table}[h!]
\centering
\caption{Accuracy and Classification Metrics on SST-2 Dataset}
\label{tab:acc_sst-2}
\begin{tabular}{lcccc}
\hline
$N_{comp}$ & Acc (\%) & Precision & F1 Score & AUC \\
\hline
\multicolumn{5}{c}{\textbf{Vanilla Transformer}} \\
-- & 90.85 & 0.909 & 0.909 & 0.959 \\
\hline
\multicolumn{5}{c}{\textbf{Differential Transformer}} \\
2 & 91.53 & 0.915 & 0.915 & 0.971 \\
\hline
\multicolumn{5}{c}{\textbf{NDT [0,1]}} \\
2 & 91.74 & 0.917 & 0.917 & 0.964 \\
3 & 91.85 & 0.919 & 0.918 & 0.962 \\
4 & 91.65 & 0.916 & 0.916 & 0.963 \\
\hline
\multicolumn{5}{c}{\textbf{NDT [-1,1]}} \\
2 & 91.83 & 0.919 & 0.918 & 0.965 \\
3 & 91.67 & 0.917 & 0.918 & 0.962 \\
4 & 91.65 & 0.916 & 0.916 & 0.963 \\
\hline
\multicolumn{5}{c}{\textbf{NDT [0,$\infty$)}} \\
2 & 91.40 & 0.914 & 0.914 & 0.965 \\
3 & 91.52 & 0.915 & 0.915 & 0.964 \\
4 & 91.06 & 0.911 & 0.910 & 0.985 \\
\hline
\multicolumn{5}{c}{\textbf{NDT ($-\infty$,$\infty$)}} \\
2 & 91.44 & 0.914 & 0.914 & 0.965 \\
3 & 91.70 & 0.917 & 0.917 & 0.964 \\
4 & 91.09 & 0.911 & 0.911 & 0.964 \\
\hline
\end{tabular}
\end{table}

\subsubsection{Lambda Value Analysis}

\textit{SST-2} $\lambda$ values (Table~\ref{tab:lambda_sst-2}) exhibit patterns largely consistent with IMDB, with notable magnitude differences. The \textbf{NDT [0,1]} configuration maintains balanced positive contributions with the familiar decreasing trend (0.1554~$\rightarrow$~0.1342 in Layer~1), with notably higher Layer~2 values (up to 0.2032) indicating stronger specialized contributions in deeper layers. Distinctively, the \textbf{NDT [-1,1]} variant shows dramatically larger negative magnitudes in Layer~1 (-0.4230 vs. IMDB's -0.2001), but transitions to mixed or positive values in Layer~2 (0.0600 for 2 components). This layer wise shift from strong negative to near zero/positive values is unique to SST-2, suggesting different suppression strategies at different depths for handling shorter, syntactically complex sentences. The unbounded configurations show higher overall magnitudes (up to 0.5294) compared to IMDB, with consistent hierarchical decrease across components.

These patterns suggest that SST-2’s linguistic characteristics (shorter text length and greater syntactic complexity) require different component weighting strategies, particularly in how subtractive mechanisms are employed across network depths.

\begin{table}[h!]
\centering
\caption{Learned Lambda Values for the SST-2 Dataset}
\label{tab:lambda_sst-2}
\begin{tabular}{lcc}
\hline
$N_{comp}$ & \makecell{Layer 1\\$\lambda$ Values} & \makecell{Layer 2\\$\lambda$ Values} \\
\hline
\multicolumn{3}{c}{\textbf{NDT [0,1]}} \\
2 & (0.1554)                 & (0.2032) \\
3 & (0.1518, 0.1498)         & (0.1990, 0.1746) \\
4 & (0.1508, 0.1428, 0.1342) & (0.1987, 0.1721, 0.1591) \\
\hline
\multicolumn{3}{c}{\textbf{NDT [-1,1]}} \\
2 & (-0.4230)                   & (0.0600) \\
3 & (-0.3575, -0.3355)          & (-0.0312, -0.0905) \\
4 & (-0.3115, -0.2488, -0.2250) & (-0.0358, -0.0777, -0.0386) \\
\hline
\multicolumn{3}{c}{\textbf{NDT [0,$\infty$)}} \\
2 & (0.2356)                  & (0.4322) \\
3 & (0.2730,  0.2587)         & (0.4027, 0.2588) \\
4 & (0.2218, 0.1845, 0.1562) & (0.3829, 0.2801, 0.2539) \\
\hline
\multicolumn{3}{c}{\textbf{NDT ($-\infty$,$\infty$)}} \\
2 & (0.3523)                 & (0.5282) \\
3 & (0.3111, 0.3014)         & (0.5294, 0.4174) \\
4 & (0.2363, 0.1765, 0.2183) & (0.4619, 0.4223, 0.3746) \\
\hline
\end{tabular}
\end{table}

\subsection{YELP Reviews Results}

\subsubsection{Overall Performance Comparison}

The \textit{YELP Reviews} dataset presents a markedly different performance landscape compared to IMDB and SST-2, reflecting the increased complexity of 5 class sentiment classification (Table~\ref{tab:acc_yelp}). Surprisingly, the \textbf{Vanilla Transformer} (60.73\%) slightly outperforms the \textbf{Differential Transformer (DT)} (60.50\%), contradicting the trend observed in binary classification tasks. Among NDT variants, the best performance is achieved by \textbf{NDT [-1,1]} with 3 components at \textbf{60.82\%}, representing a modest improvement of \textbf{+0.09\%} over Vanilla but \textbf{+0.32\%} over DT.

This behavior contrasts with IMDB and SST-2, where the [0,1] constraint dominated, though the performance differences between constraint types are minimal (within 0.6\%). Performance across all NDT variants remains tightly clustered between 60.23\% and 60.82\%, with substantially less variation compared to the binary datasets, indicating that the fundamental challenge of fine grained sentiment classification limits the impact of architectural innovations.

Unlike the binary datasets where clear winners emerged, YELP shows no consistent pattern favoring specific component counts, with competitive results observed at 2, 3, and 4 components. The precision and F1 scores follow patterns similar to accuracy, while AUC values remain consistently high (0.890 - 0.893), suggesting that the models maintain good ranking ability even when exact class predictions prove challenging.

These results indicate that while the \textit{ConPlex framework} remains competitive, the benefits of specialized attention perspectives are less pronounced in multi class scenarios where class boundaries are inherently more ambiguous.

\begin{table}[h!]
\centering
\caption{Accuracy and Classification Metrics on YELP Reviews Dataset}
\label{tab:acc_yelp}
\begin{tabular}{lcccc}
\hline
$N_{comp}$ & Acc (\%) & Precision & F1 Score & AUC \\
\hline
\multicolumn{5}{c}{\textbf{Vanilla Transformer}} \\
-- & 60.73 & 0.611 & 0.608 & 0.892 \\
\hline
\multicolumn{5}{c}{\textbf{Differential Transformer}} \\
2 & 60.50 & 0.605 & 0.603 & 0.891 \\
\hline
\multicolumn{5}{c}{\textbf{NDT [0,1]}} \\
2 & 60.36 & 0.603 & 0.603 & 0.891 \\
3 & 60.46 & 0.611 & 0.606 & 0.890 \\
4 & 60.48 & 0.597 & 0.598 & 0.891 \\
\hline
\multicolumn{5}{c}{\textbf{NDT [-1,1]}} \\
2 & 60.23 & 0.597 & 0.598 & 0.890 \\
3 & 60.82 & 0.608 & 0.608 & 0.893 \\
4 & 60.56 & 0.606 & 0.605 & 0.892 \\
\hline
\multicolumn{5}{c}{\textbf{NDT [0,$\infty$)}} \\
2 & 60.40 & 0.602 & 0.602 & 0.891 \\
3 & 60.42 & 0.612 & 0.606 & 0.890 \\
4 & 60.45 & 0.596 & 0.597 & 0.891 \\
\hline
\multicolumn{5}{c}{\textbf{NDT ($-\infty$,$\infty$)}} \\
2 & 60.39 & 0.602 & 0.603 & 0.892 \\
3 & 60.38 & 0.611 & 0.606 & 0.890 \\
4 & 60.45 & 0.597 & 0.597 & 0.891 \\
\hline
\end{tabular}
\end{table}

\subsubsection{Lambda Value Analysis}

\textit{YELP} $\lambda$ values (Table~\ref{tab:lambda_yelp}) closely mirror IMDB patterns despite dramatically different classification performance, suggesting consistent component weighting strategies across binary and multi class tasks. The \textbf{NDT [0,1]} configuration maintains the characteristic positive, decreasing trend (0.1568~$\rightarrow$~0.1410 in Layer~1) with increased Layer~2 magnitudes (up to 0.1926). The \textbf{NDT [-1,1]} variant shows exclusively negative values (-0.2003 to -0.0468) matching IMDB's magnitude range, while unbounded configurations demonstrate magnitudes (up to 0.5085) similar to SST-2. The hierarchical decrease across components persists in all configurations. Despite YELP's tight performance clustering, the well structured $\lambda$ patterns suggest the challenge lies in fundamental class ambiguity rather than inappropriate component weighting, providing evidence that ConPlex successfully enables specialization even when downstream tasks limit performance ceilings.

\begin{table}[h!]
\centering
\caption{Learned Lambda Values for the YELP Reviews Dataset}
\label{tab:lambda_yelp}
\begin{tabular}{lcc}
\hline
$N_{comp}$ & \makecell{Layer 1\\$\lambda$ Values} & \makecell{Layer 2\\$\lambda$ Values} \\
\hline
\multicolumn{3}{c}{\textbf{NDT [0,1]}} \\
2 & (0.1568)                 & (0.1926) \\
3 & (0.1520, 0.1462)         & (0.1907, 0.1635) \\
4 & (0.1527, 0.1463, 0.1410) & (0.1894, 0.1634, 0.1558) \\
\hline
\multicolumn{3}{c}{\textbf{NDT [-1,1]}} \\
2 & (-0.2003)                   & (-0.0468) \\
3 & (-0.2121, -0.1778)          & (-0.0680, -0.0850) \\
4 & (-0.2031, -0.1703, -0.1464) & (-0.0525, -0.0631, -0.0596) \\
\hline
\multicolumn{3}{c}{\textbf{NDT [0,$\infty$)}} \\
2 & (0.2429)                  & (0.3972) \\
3 & (0.2120, 0.1943)         & (0.3909, 0.2320) \\
4 & (0.2160, 0.1827, 0.1722) & (0.3719, 0.2280, 0.1994) \\
\hline
\multicolumn{3}{c}{\textbf{NDT ($-\infty$,$\infty$)}} \\
2 & (0.2997)                 & (0.5085) \\
3 & (0.2038, 0.2069)         & (0.4203, 0.2543) \\
4 & (0.2151, 0.1943, 0.1978) & (0.3621, 0.2436, 0.2380) \\
\hline
\end{tabular}
\end{table}

\subsection{Twitter Financial News Results}

\subsubsection{Overall Performance Comparison}

The \textit{Twitter Financial News} dataset, representing a 3 class sentiment classification task, demonstrates performance patterns more similar to the binary datasets (IMDB and SST-2) than to the 5 class YELP dataset (Table~\ref{tab:acc_twitfin}). The NDT framework achieves strong improvements over baseline architectures, with \textbf{NDT ($-\infty,\infty$)} with 4 components reaching the highest accuracy of \textbf{83.33\%}, representing improvements of \textbf{+1.34\%} over the Vanilla Transformer and \textbf{+1.00\%} over DT.

Notably, this dataset exhibits a clear preference for 4 component architectures, with the top three performing models all using 4 components across different constraint configurations: ($-\infty,\infty$) at $83.33\%$, [0,1] at $83.25\%$, and [$-1,1$] at $83.08\%$. This contrasts with IMDB's preference for 2 components and SST-2's preference for 3 components, suggesting that the financial domain's specialized vocabulary and distinct sentiment expressions may benefit from additional specialized attention perspectives.

The strong performance of the purely positive [0,1] constraint ($83.25\%$ with 4 components) reaffirms that constructive integration remains effective even in multi class scenarios, contradicting any assumption that subtractive mechanisms are inherently superior for handling multiple sentiment categories. Performance spread across NDT variants is moderate ($80.90\%$ to $83.33\%$), wider than YELP but narrower than IMDB, indicating reasonable architectural sensitivity. The consistently high AUC scores ($0.934$ - $0.948$), combined with strong accuracy, demonstrate robust discriminative capability across all three sentiment classes.

Overall, 8 out of 12 NDT variants surpass the DT baseline, reinforcing the generalizability of the \textit{ConPlex framework} across diverse sentiment analysis contexts.

\begin{table}[h!]
\centering
\caption{Accuracy and Classification Metrics on Twitter Financial News Dataset}
\label{tab:acc_twitfin}
\begin{tabular}{lcccc}
\hline
$N_{comp}$ & Acc (\%) & Precision & F1 Score & AUC \\
\hline
\multicolumn{5}{c}{\textbf{Vanilla Transformer}} \\
-- & 81.99 & 0.817 & 0.817 & 0.936 \\
\hline
\multicolumn{5}{c}{\textbf{Differential Transformer}} \\
2 & 82.33 & 0.819 & 0.821 & 0.945 \\
\hline
\multicolumn{5}{c}{\textbf{NDT [0,1]}} \\
2 & 80.90 & 0.808 & 0.808 & 0.940 \\
3 & 82.50 & 0.826 & 0.824 & 0.941 \\
4 & 83.25 & 0.828 & 0.829 & 0.948 \\
\hline
\multicolumn{5}{c}{\textbf{NDT [-1,1]}} \\
2 & 82.66 & 0.827 & 0.827 & 0.937 \\
3 & 82.33 & 0.823 & 0.822 & 0.942 \\
4 & 83.08 & 0.830 & 0.830 & 0.943 \\
\hline
\multicolumn{5}{c}{\textbf{NDT [0,$\infty$)}} \\
2 & 81.49 & 0.814 & 0.812 & 0.939 \\
3 & 82.75 & 0.827 & 0.827 & 0.934 \\
4 & 82.41 & 0.824 & 0.824 & 0.936 \\
\hline
\multicolumn{5}{c}{\textbf{NDT ($-\infty$,$\infty$)}} \\
2 & 82.41 & 0.825 & 0.824 & 0.939 \\
3 & 82.33 & 0.822 & 0.822 & 0.939 \\
4 & 83.33 & 0.829 & 0.830 & 0.948 \\
\hline
\end{tabular}
\end{table}

\subsubsection{Lambda Value Analysis}

\textit{Twitter Financial News} $\lambda$ values (Table~\ref{tab:lambda_twitfin}) remarkably mirror IMDB patterns despite domain differences and 3 class classification. The \textbf{NDT [0,1]} configuration maintains balanced positive contributions with the characteristic decreasing trend (0.1522~$\rightarrow$~0.1372 in Layer~1) and modest Layer~2 increases (up to 0.1841), matching IMDB's magnitude range. The \textbf{NDT [-1,1]} variant learns exclusively negative values (-0.2053 to -0.0593) closely matching IMDB, with stable negative values across both layers (unlike SST-2's layer wise transitions). The unbounded configurations show moderate magnitudes (up to 0.3741), falling between IMDB and SST-2/YELP values. Notably, the ($-\infty$,$\infty$) configuration shows unusually high Layer~1 values (0.3261), aligning with this configuration achieving best performance (83.33\%). The overall stability across Twitter, IMDB, and YELP (contrasting SST-2) suggests dataset size and syntactic complexity influence component weighting more than output class count or domain characteristics.

\begin{table}[h!]
\centering
\caption{Learned Lambda Values for the Twitter Financial News Dataset}
\label{tab:lambda_twitfin}
\begin{tabular}{lcc}
\hline
$N_{comp}$ & \makecell{Layer 1\\$\lambda$ Values} & \makecell{Layer 2\\$\lambda$ Values} \\
\hline
\multicolumn{3}{c}{\textbf{NDT [0,1]}} \\
2 & (0.1522)                 & (0.1841) \\
3 & (0.1512, 0.1444)         & (0.1839, 0.1607) \\
4 & (0.1516, 0.1441, 0.1372) & (0.1822, 0.1591, 0.1515) \\
\hline
\multicolumn{3}{c}{\textbf{NDT [-1,1]}} \\
2 & (-0.2053)                   & (-0.0593) \\
3 & (-0.2012, -0.1604)          & (-0.0661, -0.0900) \\
4 & (-0.1996, -0.1599, -0.1282) & (-0.0541, -0.0802, -0.0547) \\
\hline
\multicolumn{3}{c}{\textbf{NDT [0,$\infty$)}} \\
2 & (0.2248)                  & (0.3707) \\
3 & (0.2303, 0.2045)         & (0.3741, 0.2178) \\
4 & (0.2368, 0.1971, 0.1652) & (0.3669, 0.2167, 0.1792) \\
\hline
\multicolumn{3}{c}{\textbf{NDT ($-\infty$,$\infty$)}} \\
2 & (0.3261)                 & (0.3571) \\
3 & (0.2144, 0.2087)         & (0.3572, 0.2234) \\
4 & (0.2372, 0.2053, 0.1637) & (0.3532, 0.2015, 0.1671) \\
\hline
\end{tabular}
\end{table}

\subsection{Representation Quality Analysis}
Table~\ref{tab:rep_quality} presents the silhouette scores across all four datasets, providing insights into how well the learned embeddings cluster samples according to their sentiment classes. The silhouette score quantifies embedding quality by measuring intraclass compactness relative to interclass separation, with higher values indicating better defined clusters.

\textbf{Patterns Across Datasets:} The representation quality varies dramatically across datasets, with SST-2 demonstrating the highest silhouette scores ($0.360$ - $0.499$) and YELP showing the lowest ($0.043$ - $0.059$). This stark contrast reflects the fundamental difference between binary and fine grained multi class classification: binary sentiment distinctions create naturally separable clusters in embedding space, while 5 class ratings occupy a continuous spectrum with inherently overlapping boundaries. IMDB exhibits moderate scores ($0.261$ - $0.361$), while \textit{Twitter Financial News} shows intermediate behavior ($0.114$ - $0.276$), consistent with its 3 class structure.

\textbf{Model Comparisons:} For SST-2, the Differential Transformer achieves an exceptionally high silhouette score of $0.490$, with most NDT variants producing competitive values ($0.451$ - $0.499$). The NDT [$-1,1$] with 4 components achieves the highest score of $0.499$, demonstrating superior cluster separation. On IMDB, the DT substantially outperforms the Vanilla Transformer ($0.361$ vs.\ $0.314$), though NDT variants show more varied behavior ($0.261$ - $0.337$), with several configurations slightly underperforming DT despite achieving higher classification accuracy. This dissociation between silhouette scores and accuracy suggests that classification performance depends not solely on cluster separation but also on decision boundary learning. For \textit{Twitter Financial News}, NDT variants exhibit high variability ($0.114$ - $0.276$), with NDT [$-1,1$] and [0,$\infty$) configurations achieving the strongest scores ($0.265$ and $0.276$ with 4 and 3 components, respectively), indicating that different constraint types produce qualitatively different embedding geometries.

\textbf{YELP's Unique Challenge:} The consistently low silhouette scores across all models on YELP ($0.043$ - $0.059$) confirm that 5 class sentiment creates fundamentally ambiguous boundaries in representation space. The minimal variation between Vanilla ($0.059$), DT ($0.053$), and NDT variants ($0.043$ - $0.058$) suggests that architectural innovations cannot overcome the inherent overlap between adjacent rating categories (e.g., 3 star vs.\ 4 star reviews), where linguistic differences are often subtle or context dependent.

\textbf{Key Insight:} While higher silhouette scores generally correlate with better classification performance on binary tasks, this relationship weakens for multi class datasets. The NDT framework demonstrates competitive representation quality across all datasets, though the optimal constraint configuration for embedding separability varies by dataset characteristics, highlighting the importance of architectural flexibility in the \textsc{ConPlex} framework.

\begin{table}[h!]
\centering
\caption{Silhouette Scores across Datasets}
\label{tab:rep_quality}
\begin{tabular}{lcccc}
\hline
$N_{comp}$ & \makecell{IMDB\\Reviews} & \makecell{SST-2} & \makecell{YELP\\Reviews}    & \makecell{Twitter Financial\\News} \\
\hline
\multicolumn{5}{c}{\textbf{Vanilla Transformer}} \\
-- & 0.314 & 0.360 & 0.059 & 0.205 \\
\hline
\multicolumn{5}{c}{\textbf{Differential Transformer}} \\
2 & 0.361 & 0.490 & 0.053 & 0.177 \\
\hline
\multicolumn{5}{c}{\textbf{NDT [0,1]}} \\
2 & 0.318 & 0.467 & 0.055 & 0.114 \\
3 & 0.291 & 0.479 & 0.050 & 0.156 \\
4 & 0.303 & 0.467 & 0.057 & 0.174 \\
\hline
\multicolumn{5}{c}{\textbf{NDT [-1,1]}} \\
2 & 0.303 & 0.469 & 0.043 & 0.261 \\
3 & 0.337 & 0.469 & 0.050 & 0.175 \\
4 & 0.335 & 0.499 & 0.051 & 0.265 \\
\hline
\multicolumn{5}{c}{\textbf{NDT [0,$\infty$)}} \\
2 & 0.325 & 0.451 & 0.053 & 0.126 \\
3 & 0.293 & 0.458 & 0.058 & 0.276 \\
4 & 0.295 & 0.455 & 0.044 & 0.220 \\
\hline
\multicolumn{5}{c}{\textbf{NDT ($-\infty$,$\infty$)}} \\
2 & 0.302 & 0.475 & 0.055 & 0.201 \\
3 & 0.317 & 0.469 & 0.054 & 0.132 \\
4 & 0.261 & 0.471 & 0.051 & 0.154 \\
\hline
\end{tabular}
\end{table}

\subsection{Visualization Analysis}

\subsubsection{Confusion Matrices}
\begin{figure}[t]
    \centering
    \begin{subfigure}[b]{0.24\textwidth}
        \centering
        \includegraphics[width=\textwidth]{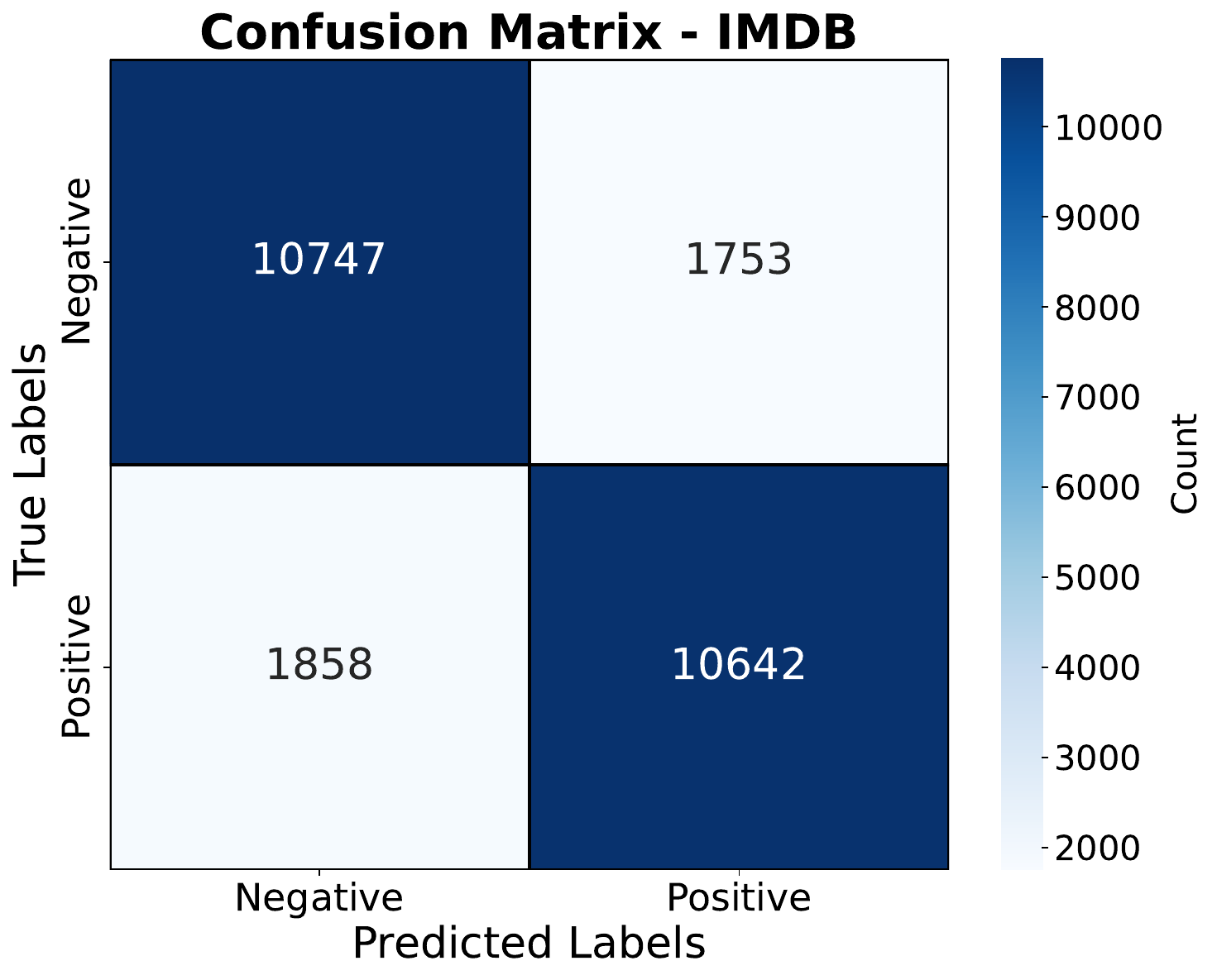}
        \caption{NDT [0,1] | comp = 2}
        \label{fig:cm_imdb}
    \end{subfigure}
    \hfill
    \begin{subfigure}[b]{0.24\textwidth}
        \centering
        \includegraphics[width=\textwidth]{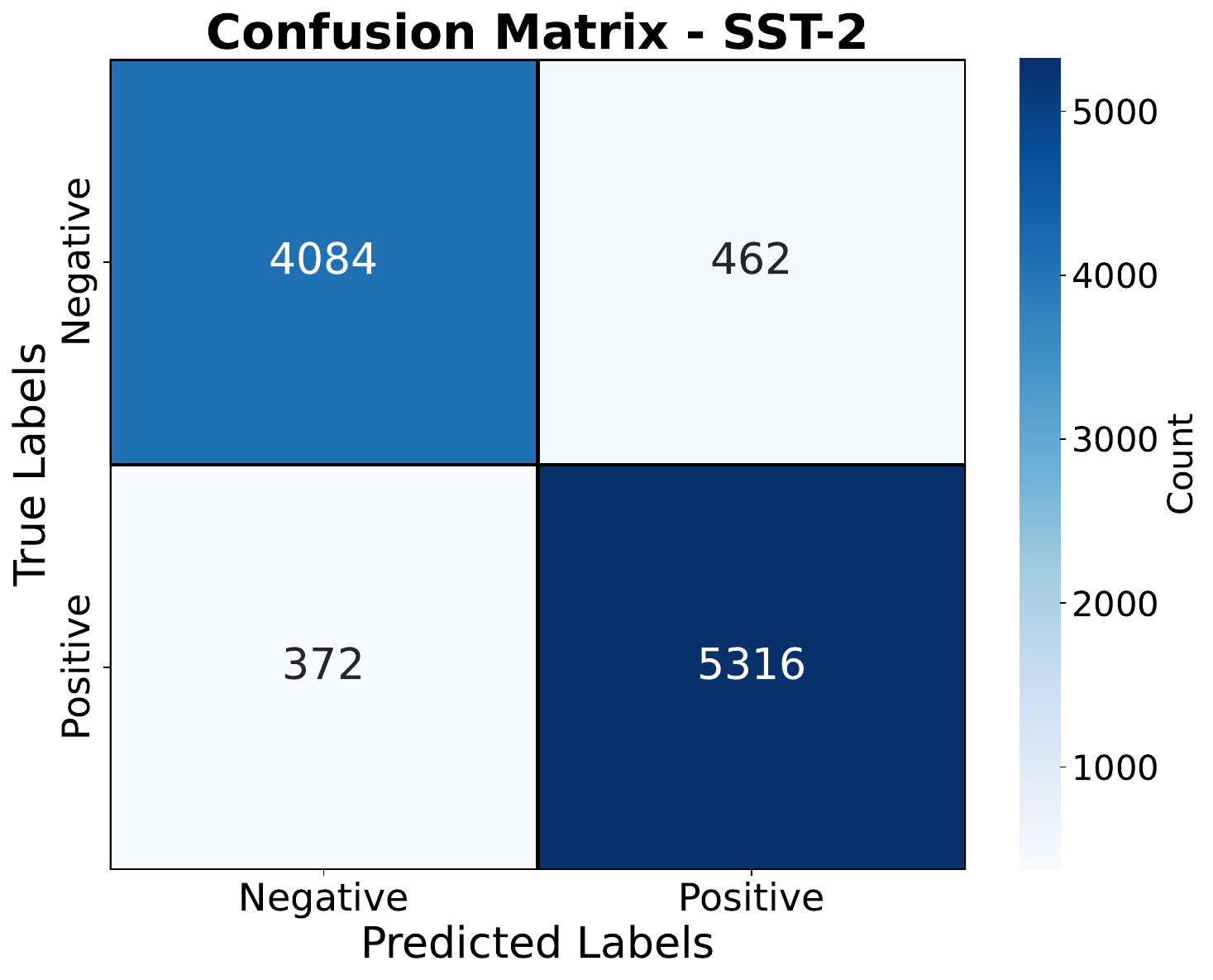}
        \caption{NDT [0,1] | comp = 3}
        \label{fig:cm_sst2}
    \end{subfigure}

    \vspace{0.25cm}

    \begin{subfigure}[b]{0.24\textwidth}
        \centering
        \includegraphics[width=\textwidth]{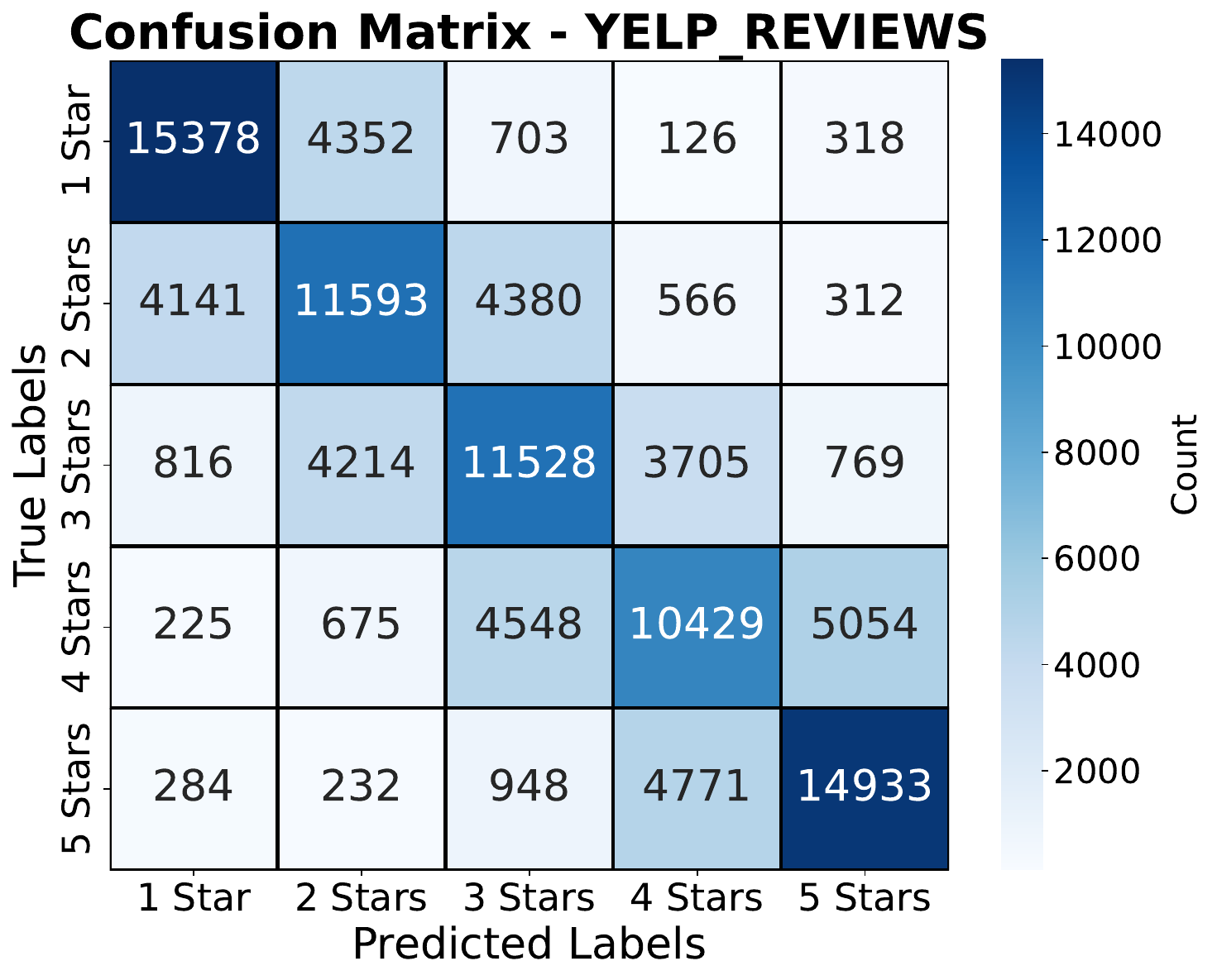}
        \caption{NDT [-1,1] | comp = 3}
        \label{fig:cm_yelp}
    \end{subfigure}
    \hfill
    \begin{subfigure}[b]{0.24\textwidth}
        \centering
        \includegraphics[width=\textwidth]{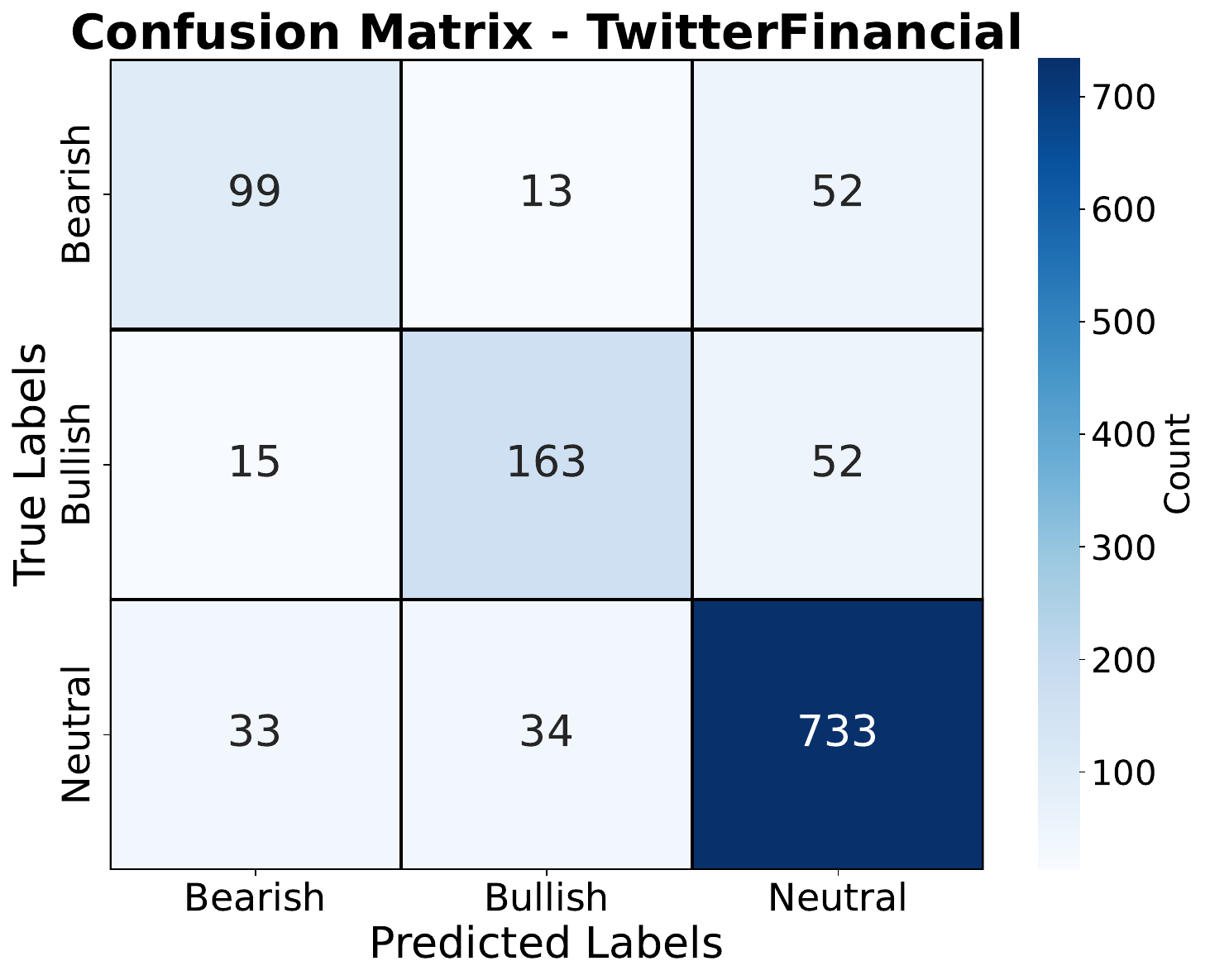}
        \caption{NDT [$-\infty$,$\infty$] | comp = 4}
        \label{fig:cm_twitfin}
    \end{subfigure}

    \caption{Confusion matrices for the best performing NDT model for each dataset. These visualizations illustrate how the model predictions are distributed across the ground truth classes.}
    \label{fig:confusion_matrices}
\end{figure}

Figure \ref{fig:confusion_matrices} represents the confusion matrices of the best performing NDT model for each dataset. 
Each subfigure corresponds to a different dataset and model configuration. These visualizations provide a clear view of the model's prediction distribution across ground truth classes, highlighting areas of correct classification as well as common misclassifications. 
Overall, the matrices demonstrate that the NDT model achieves high class separability across all datasets, with minimal confusion between classes.

\subsubsection{ROC Curves}
\begin{figure*}[ht!]
    \centering
    \begin{subfigure}[b]{0.24\textwidth}
        \centering
        \includegraphics[width=\textwidth]{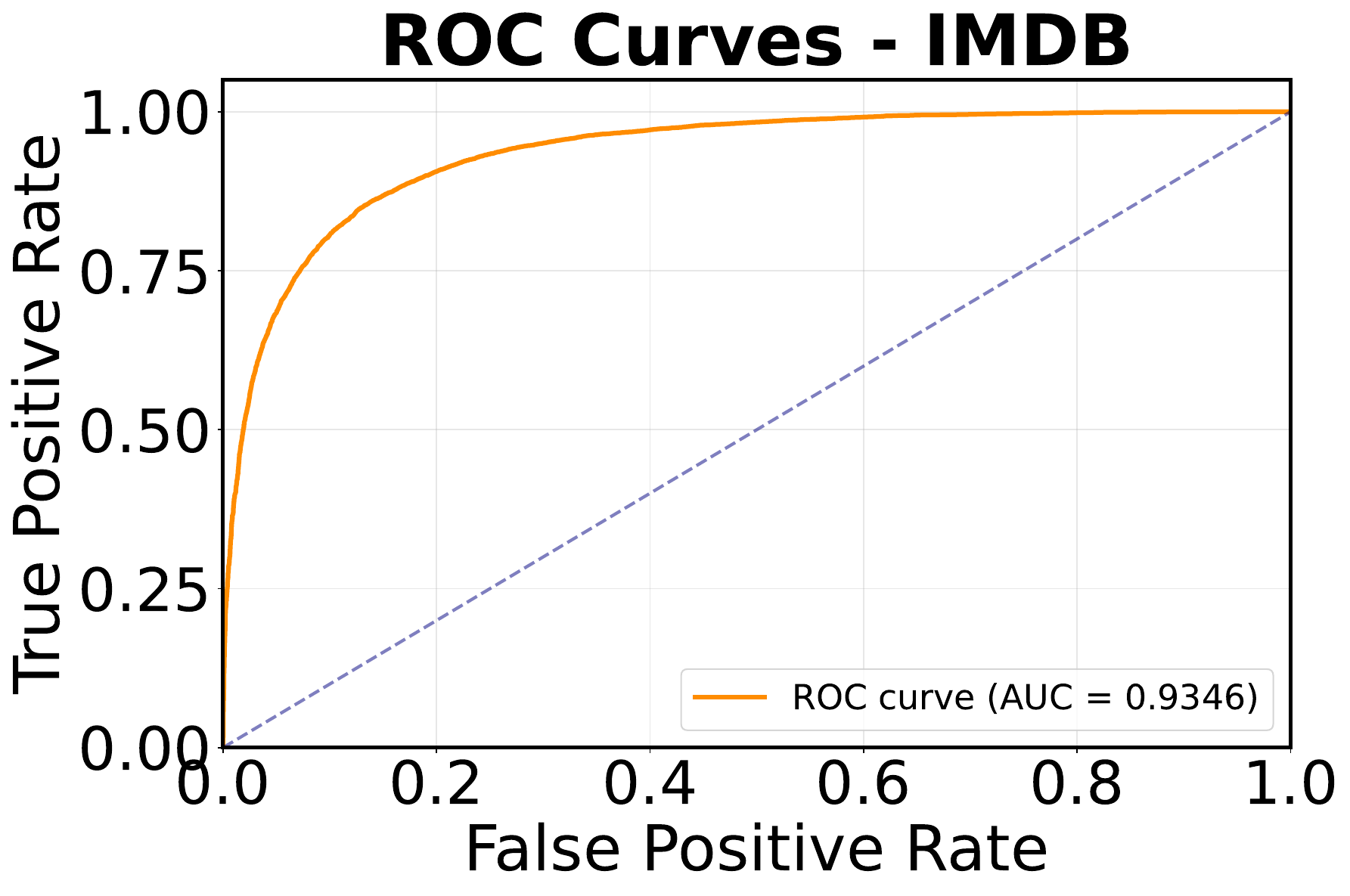}
        \caption{DT - IMDB}
        \label{fig:roc_dt_imdb}
    \end{subfigure}
    \hfill\begin{subfigure}[b]{0.24\textwidth}
        \centering
        \includegraphics[width=\textwidth]{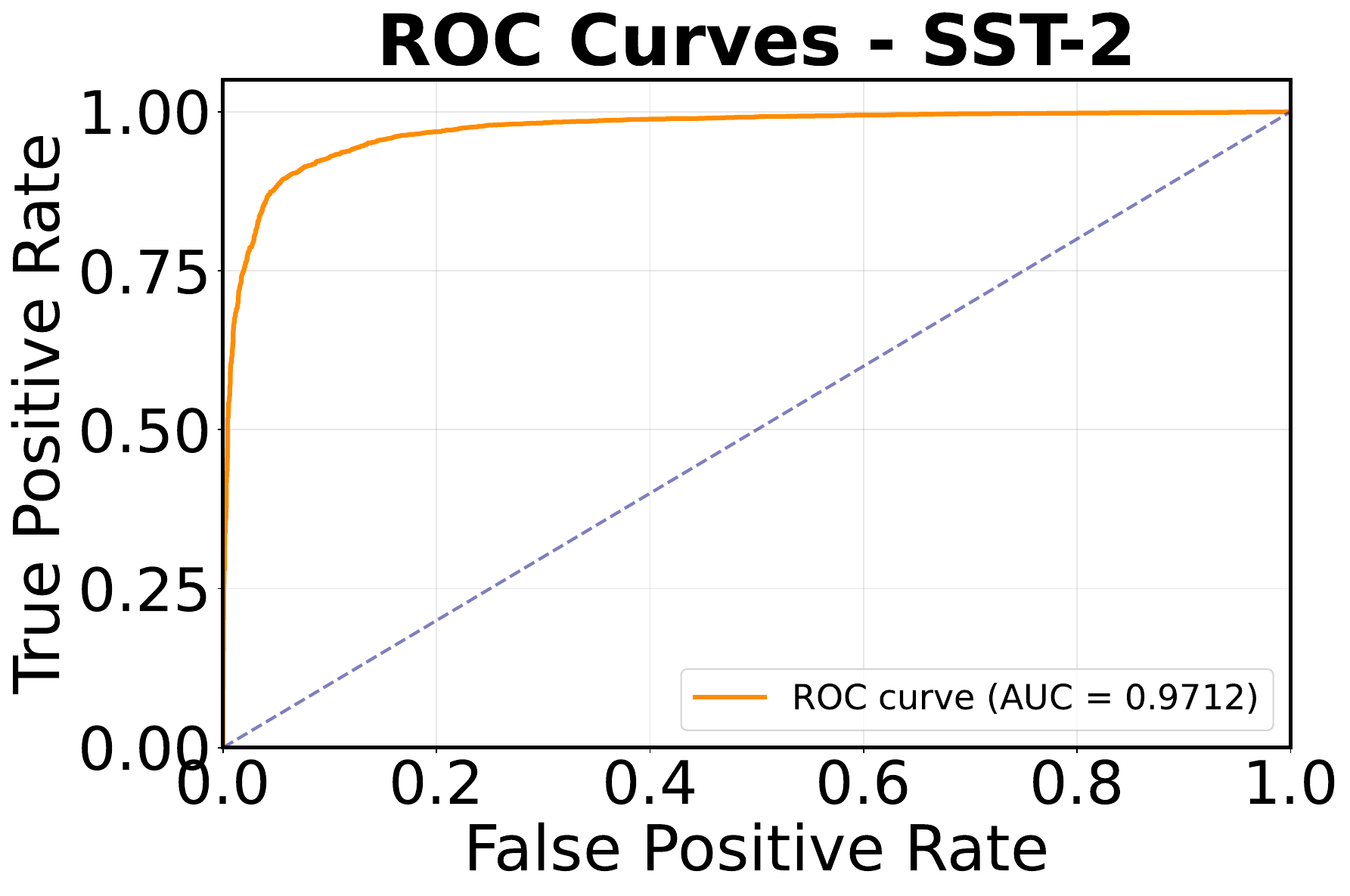}
        \caption{DT - SST-2}
        \label{fig:roc_dt_sst2}
    \end{subfigure}
    \hfill
    \begin{subfigure}[b]{0.24\textwidth}
        \centering
        \includegraphics[width=\textwidth]{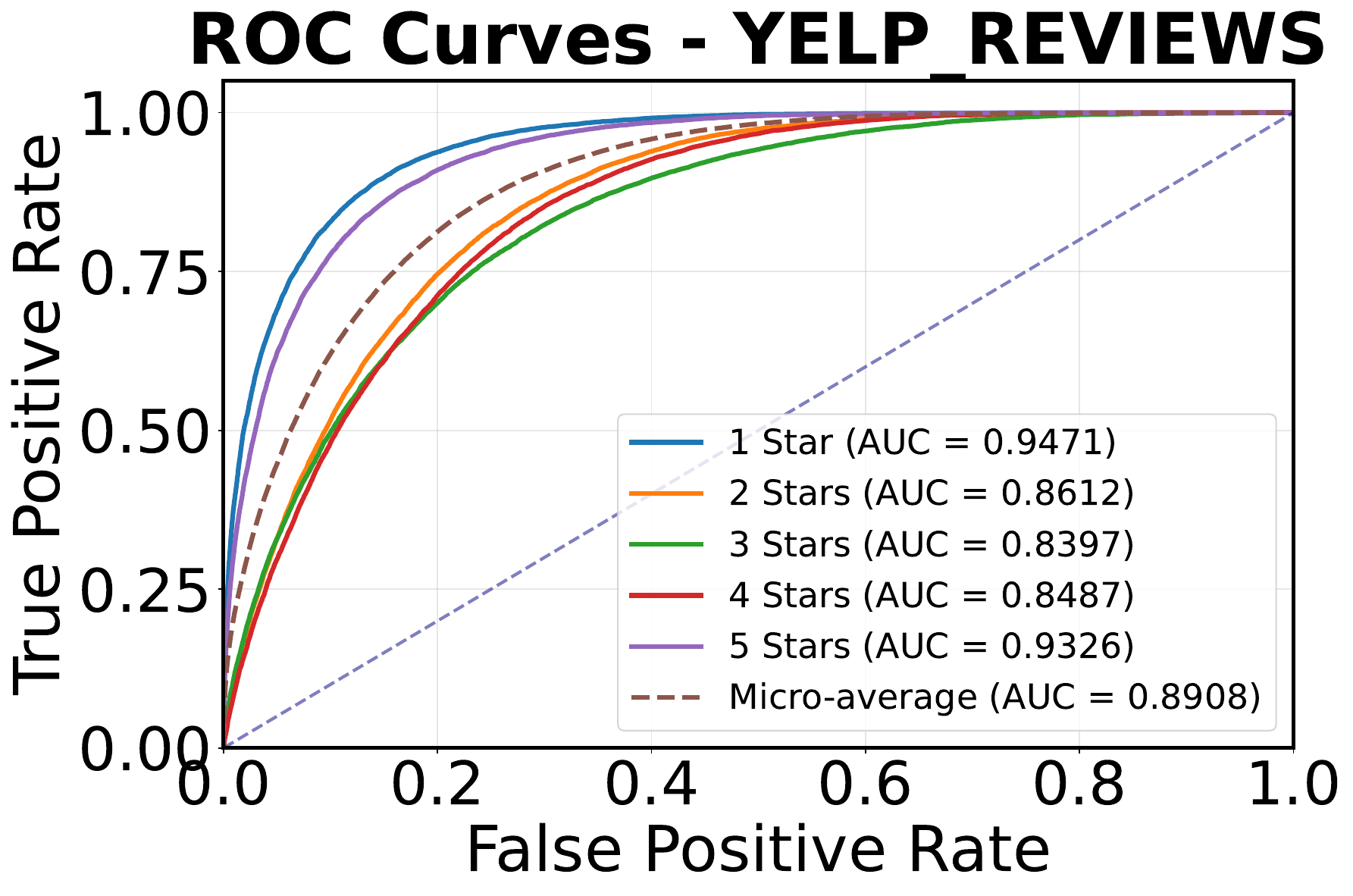}
        \caption{DT - YELP}
        \label{fig:roc_dt_yelp}
    \end{subfigure}
    \hfill
    \begin{subfigure}[b]{0.24\textwidth}
        \centering
        \includegraphics[width=\textwidth]{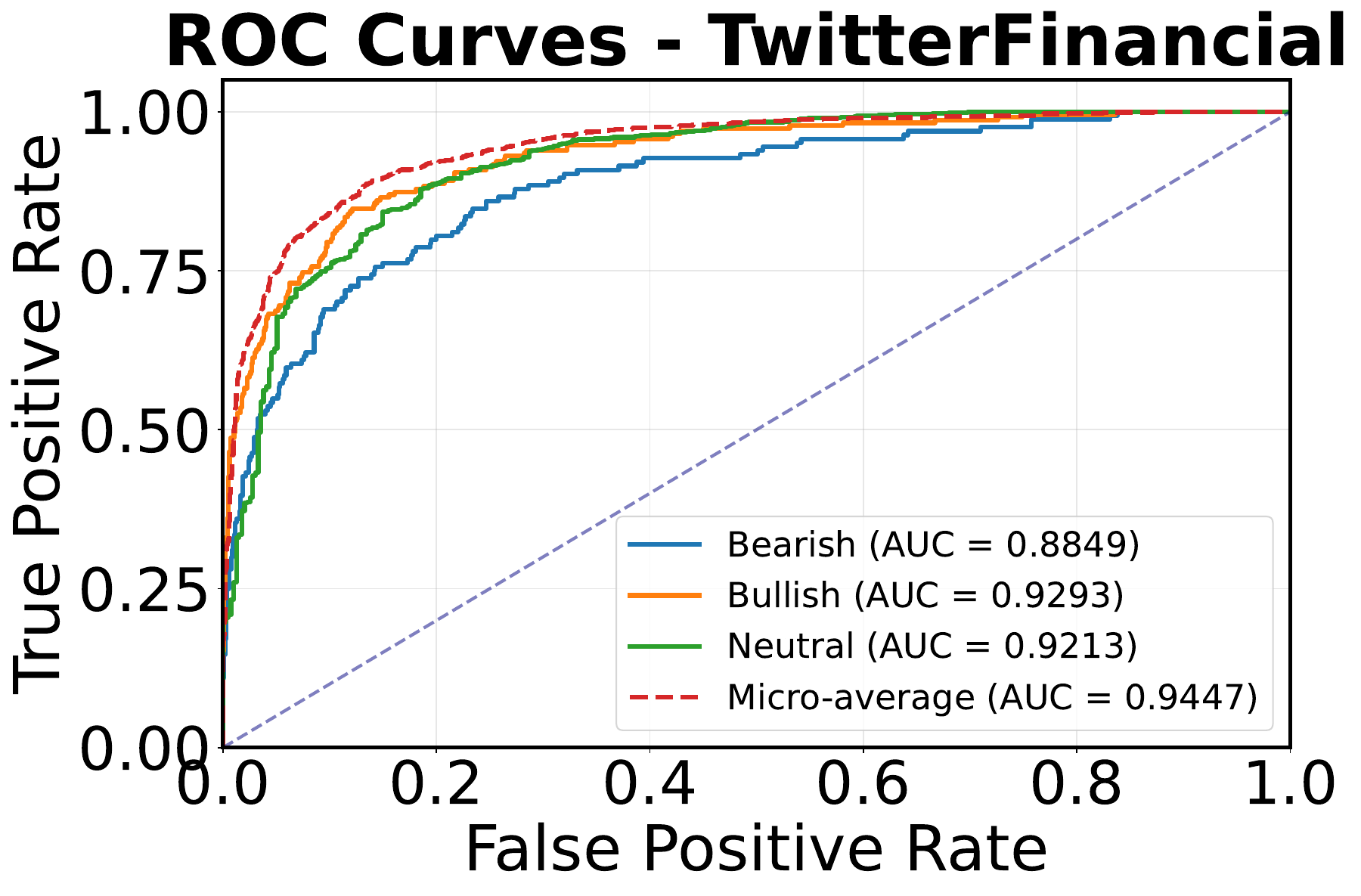}
        \caption{DT - Twitter Fin. News}
        \label{fig:roc_dt_twitfin}
    \end{subfigure}

    \vspace{0.25cm}

    \begin{subfigure}[b]{0.24\textwidth}
        \centering
        \includegraphics[width=\textwidth]{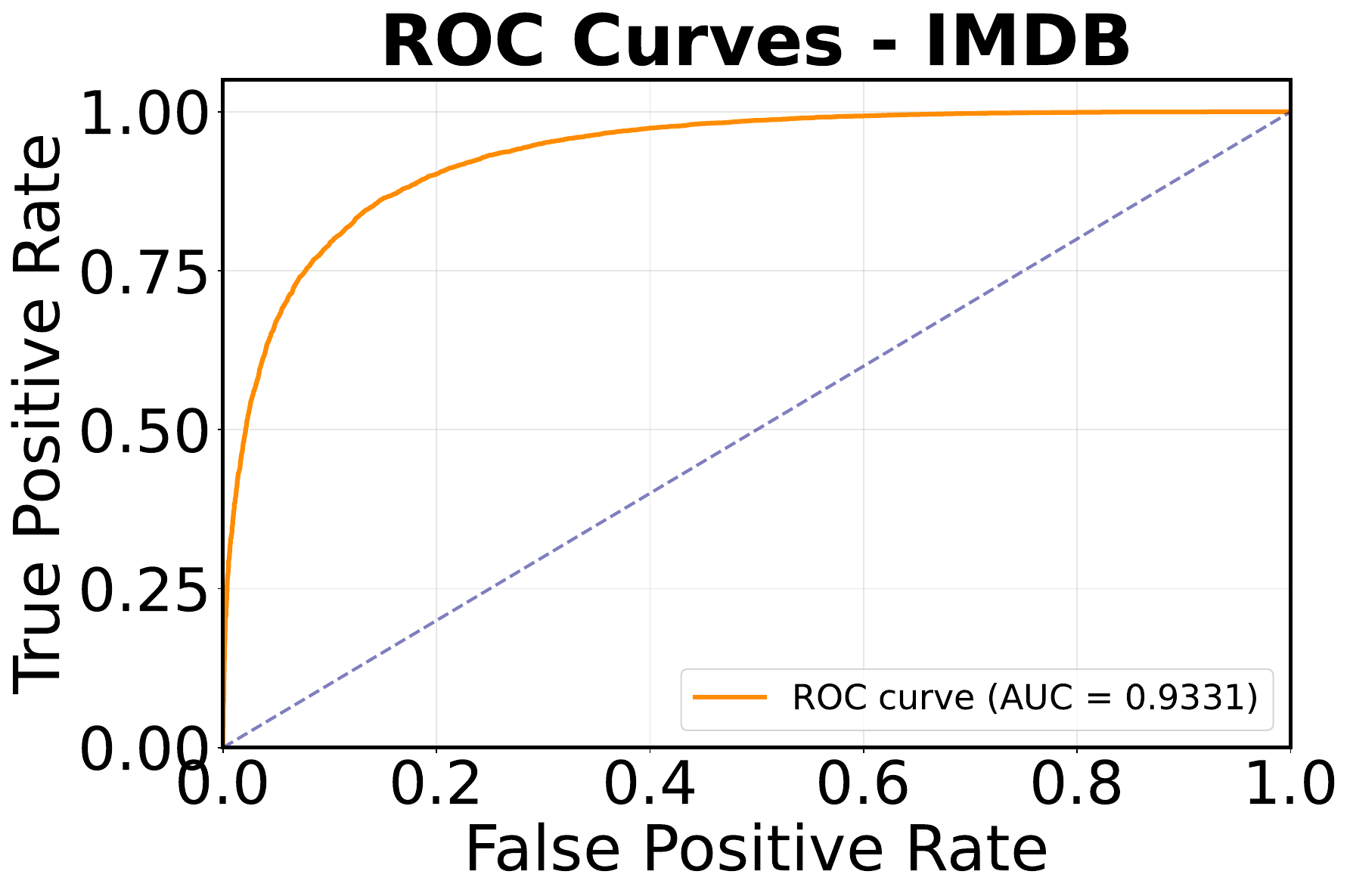}
        \caption{NDT - IMDB}
        \label{fig:roc_ndt_imdb}
    \end{subfigure}
    \hfill
    \begin{subfigure}[b]{0.24\textwidth}
        \centering
        \includegraphics[width=\textwidth]{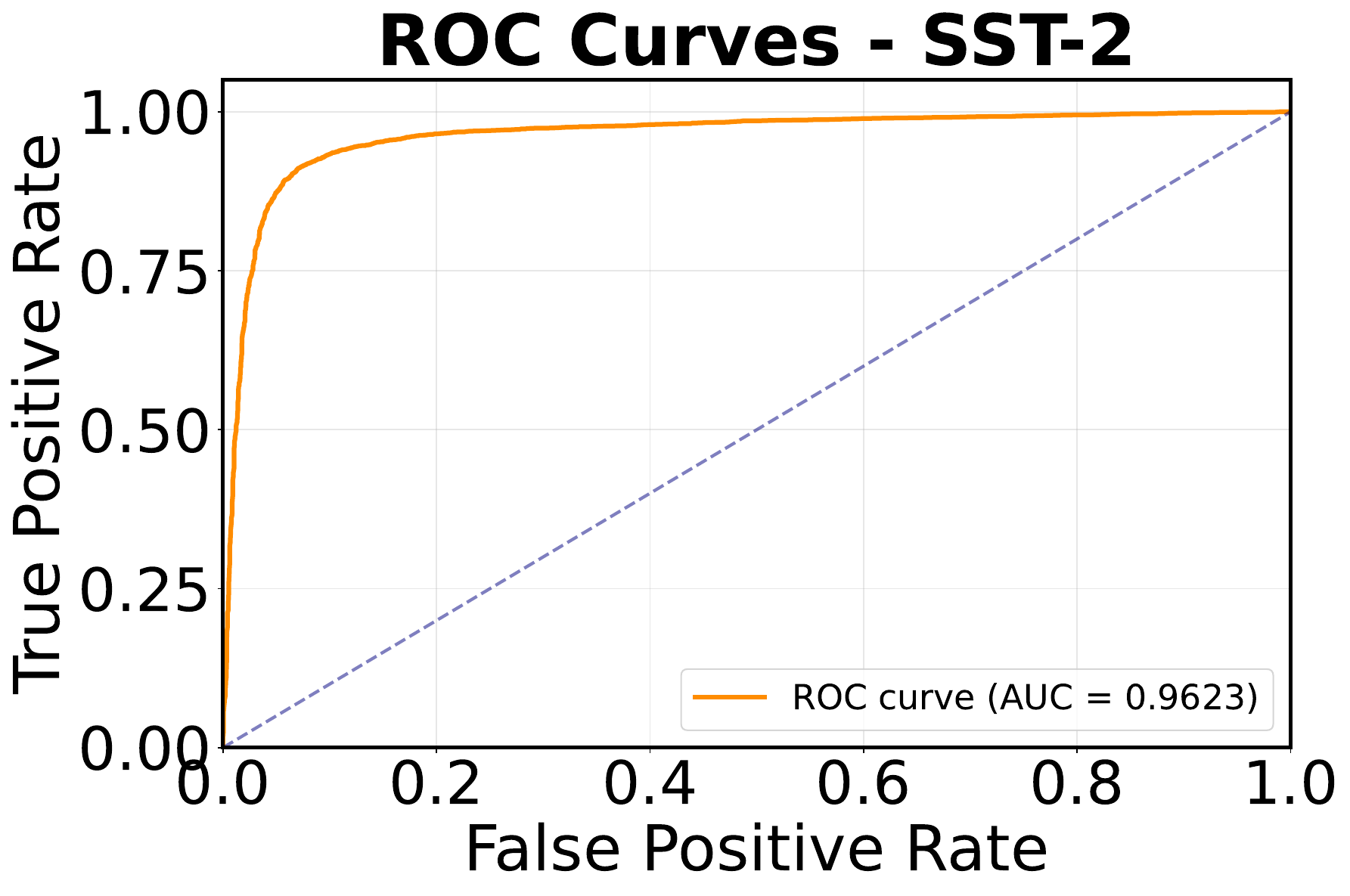}
        \caption{NDT - SST-2}
        \label{fig:roc_ndt_sst2}
    \end{subfigure}
    \hfill
    \begin{subfigure}[b]{0.24\textwidth}
        \centering
        \includegraphics[width=\textwidth]{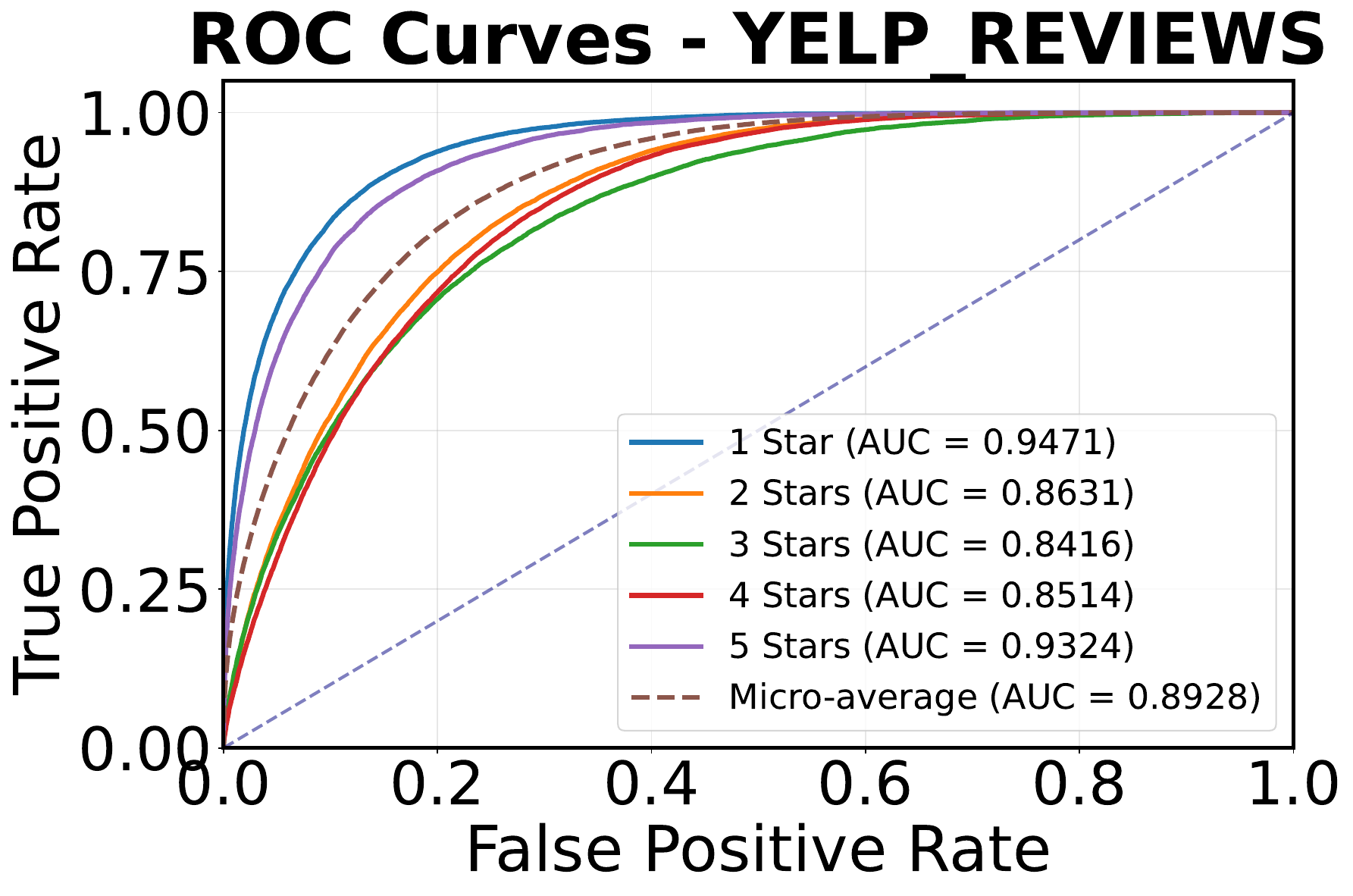}
        \caption{NDT - YELP}
        \label{fig:roc_ndt_yelp}
    \end{subfigure}
    \hfill
    \begin{subfigure}[b]{0.24\textwidth}
        \centering
        \includegraphics[width=\textwidth]{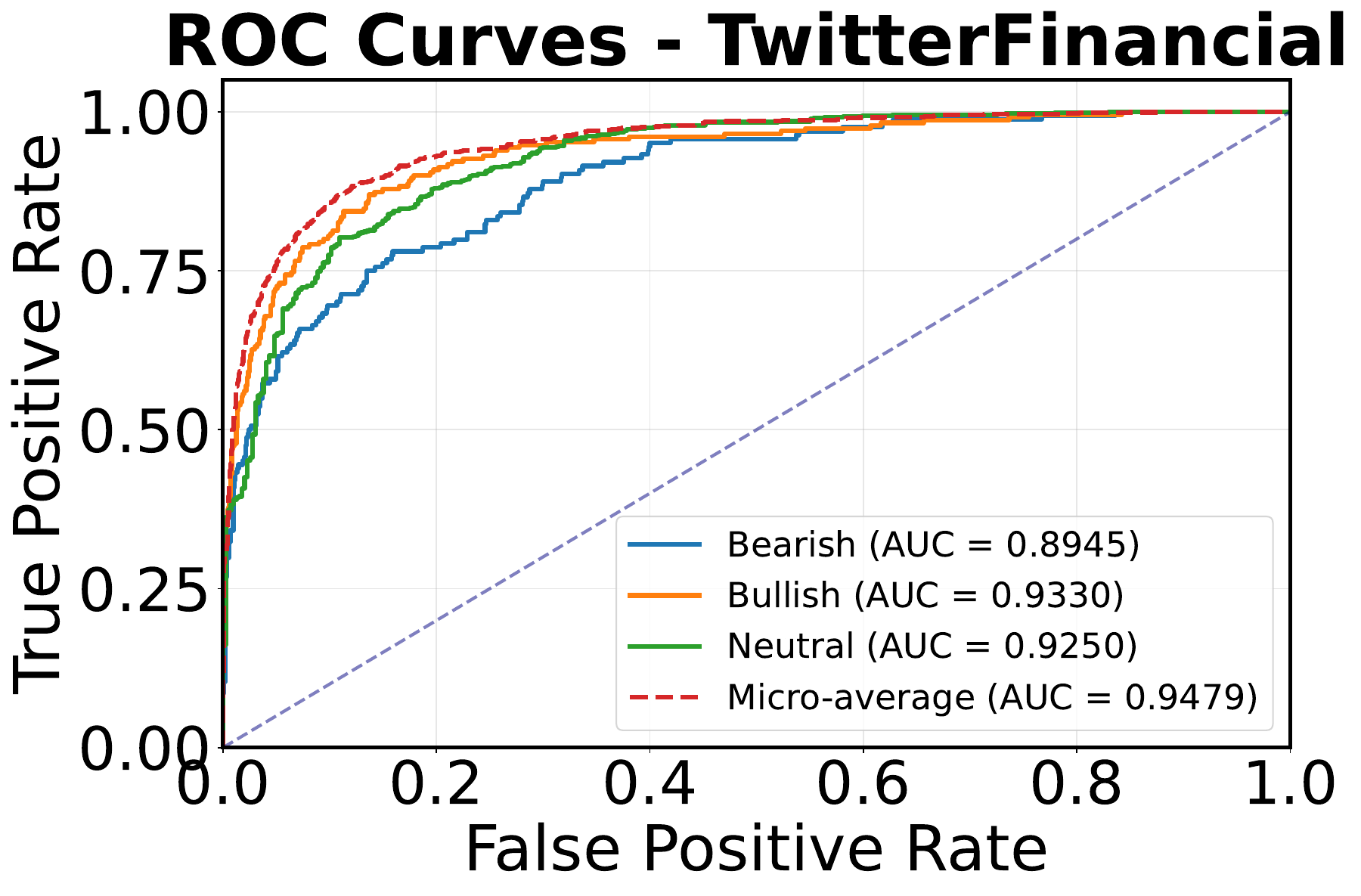}
        \caption{NDT - Twitter Fin. News}
        \label{fig:roc_ndt_twitfin}
    \end{subfigure}

    \caption{ROC curves comparing the baseline DT models (left column) with the best performing NDT models (right column) for each dataset. Each pair of columns corresponds to a dataset, illustrating the relative discriminative power of NDT and DT across settings.}
    \label{fig:roc_curves}
\end{figure*}

Figure~\ref{fig:roc_curves} presents the Receiver Operating Characteristic (ROC) curves comparing the baseline DT models (Figs.~\ref{fig:roc_dt_imdb}, \ref{fig:roc_dt_sst2}, \ref{fig:roc_dt_yelp}, \ref{fig:roc_dt_twitfin}) 
and the best performing NDT models (Figs.~\ref{fig:roc_ndt_imdb}, \ref{fig:roc_ndt_sst2}, \ref{fig:roc_ndt_yelp}, \ref{fig:roc_ndt_twitfin}) across the datasets. Each column corresponds to one dataset, with the top row showing the ROC curves for the DT models and the bottom row displaying the curves for the optimized NDT variants. The ROC curves demonstrate that while NDT achieves higher classification accuracy across datasets, DT retains higher AUC on IMDB and SST-2. NDT shows improved AUC on YELP and Twitter Financial News, indicating that the benefits of additive attention are more pronounced in multi-class settings when measured by ranking ability.

\subsubsection{t-SNE Embeddings}
\begin{figure}[ht!]
    \centering
    \begin{subfigure}[b]{0.24\textwidth}
        \centering
        \includegraphics[width=\textwidth]{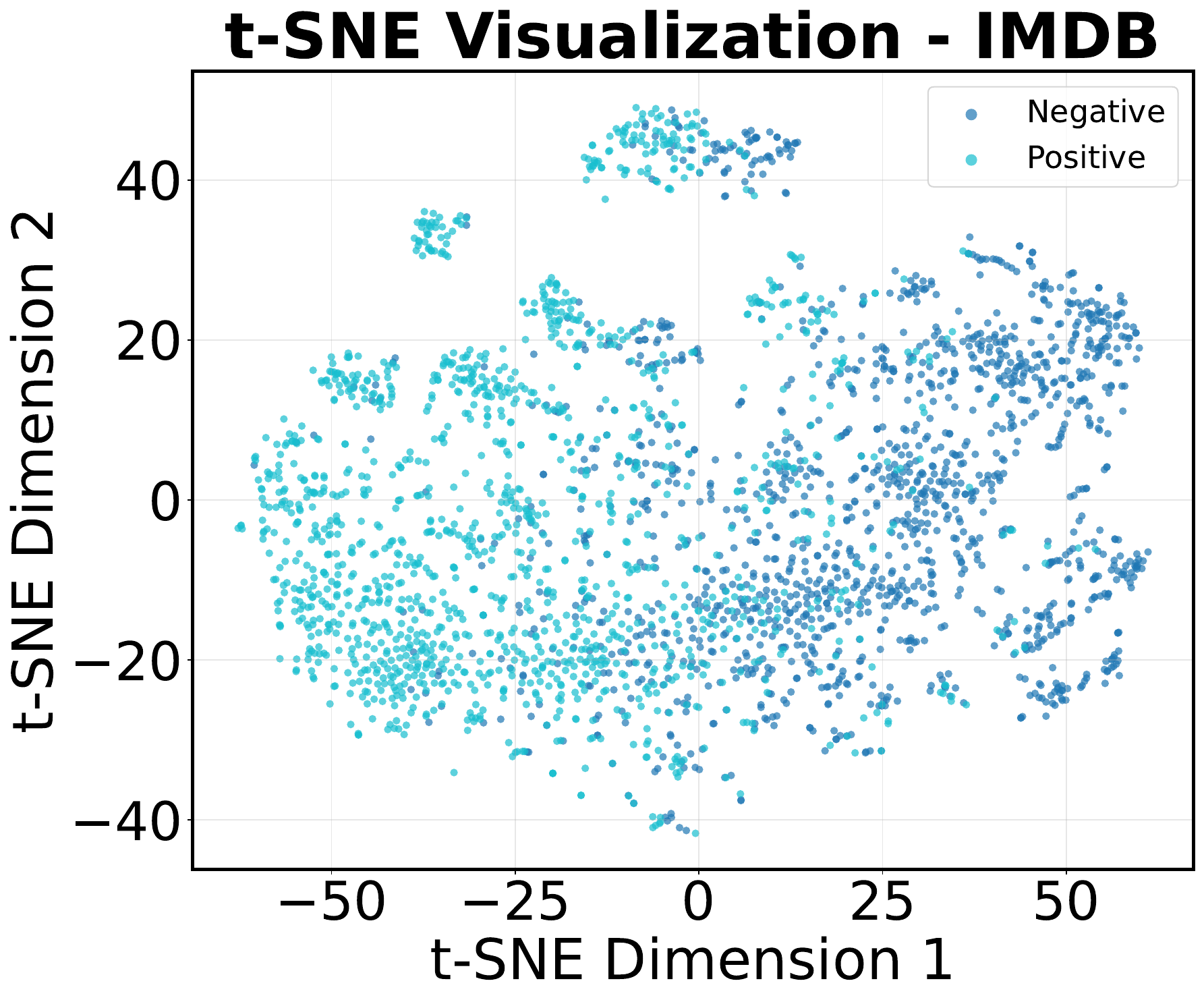}
        \caption{NDT [0,1] | comp = 2}
        \label{fig:tsne_imdb}
    \end{subfigure}
    \hfill
    \begin{subfigure}[b]{0.24\textwidth}
        \centering
        \includegraphics[width=\textwidth]{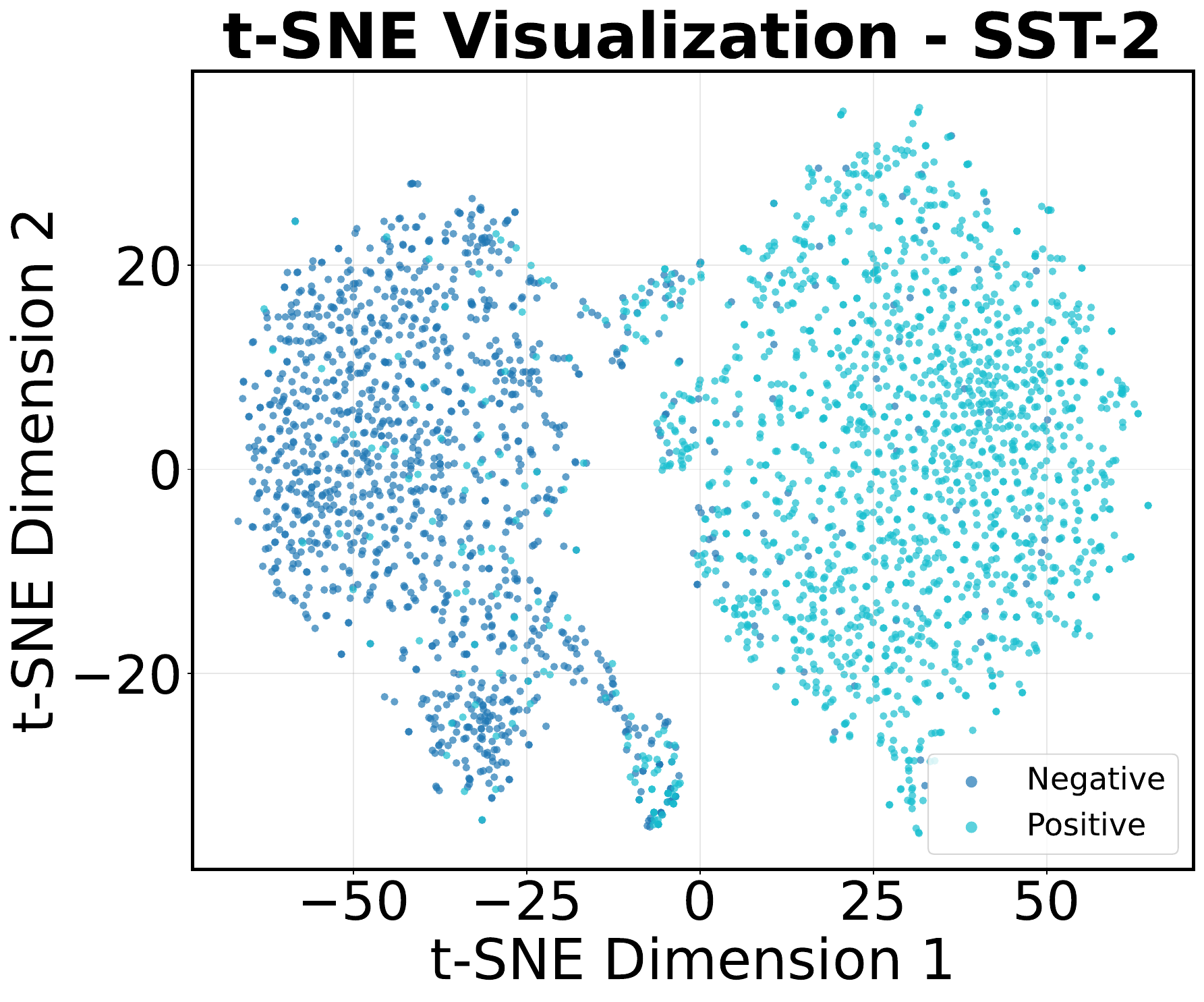}
        \caption{NDT [0,1] | comp = 3}
        \label{fig:tsne_sst2}
    \end{subfigure}

    \vspace{0.25cm}

    \begin{subfigure}[b]{0.24\textwidth}
        \centering
        \includegraphics[width=\textwidth]{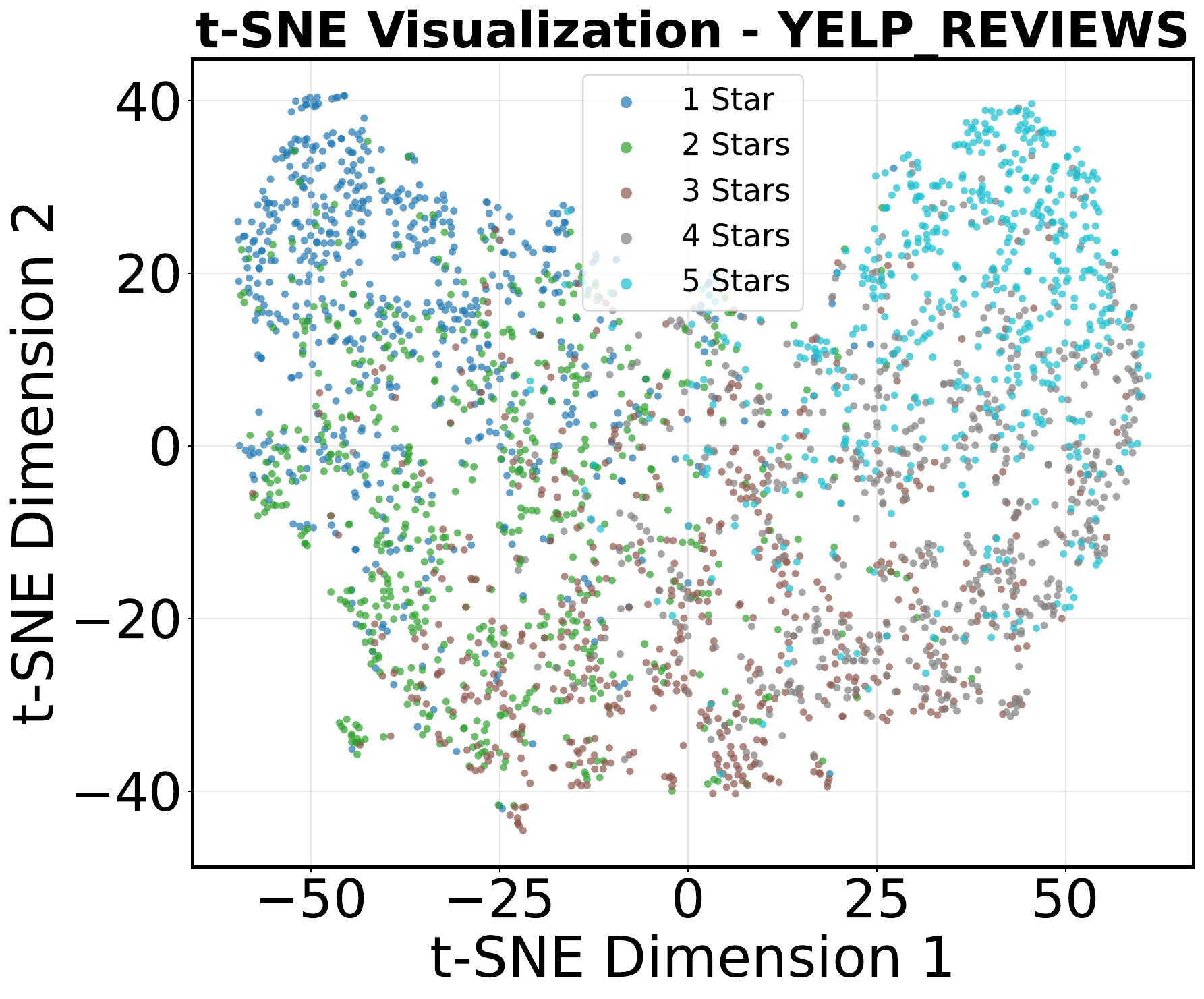}
        \caption{NDT [-1,1] | comp = 3}
        \label{fig:tsne_yelp}
    \end{subfigure}
    \hfill
    \begin{subfigure}[b]{0.24\textwidth}
        \centering
        \includegraphics[width=\textwidth]{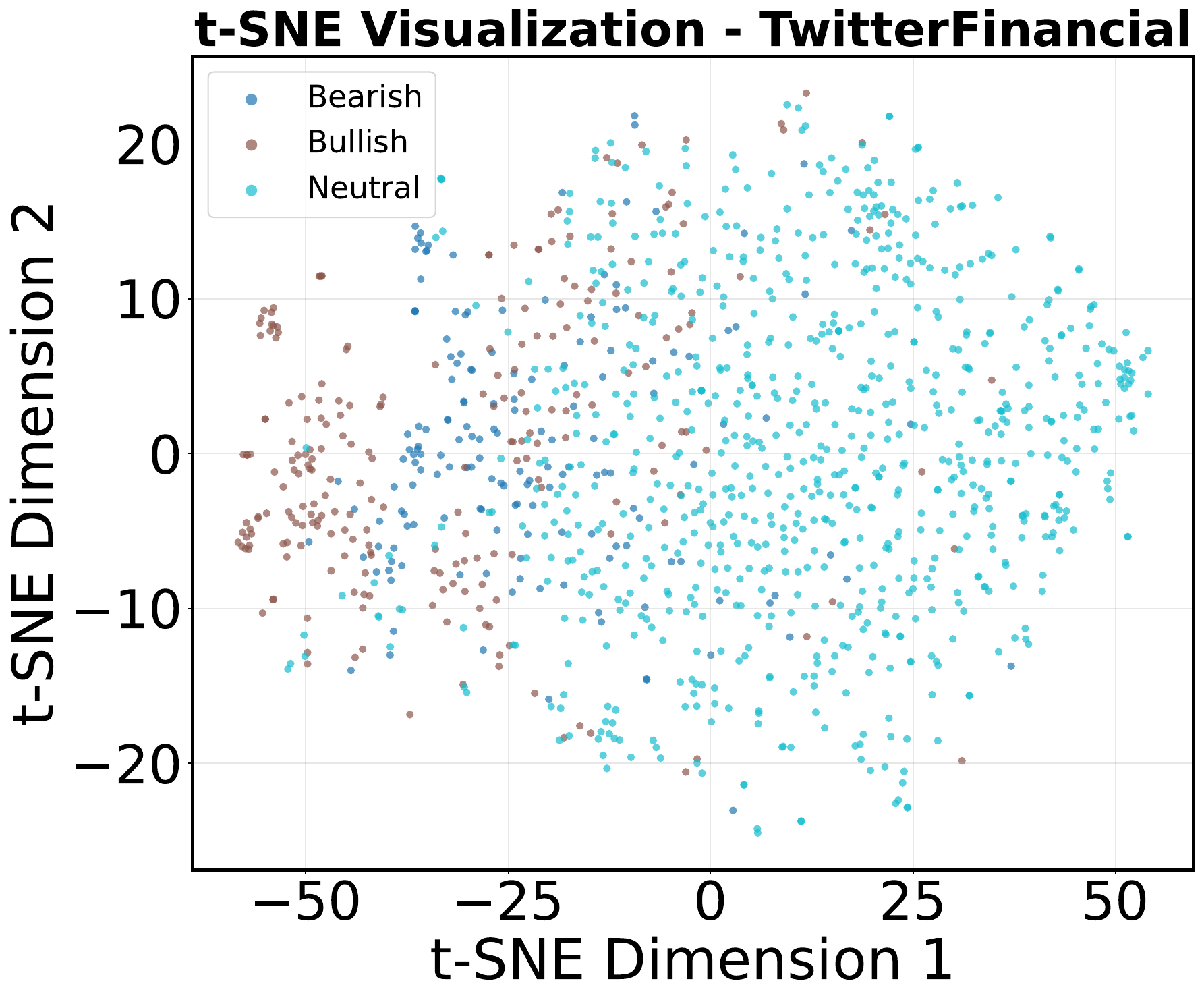}
        \caption{NDT [$-\infty$,$\infty$] | comp = 4}
        \label{fig:tsne_twitfin}
    \end{subfigure}

    \caption{t-SNE visualizations of the learned embeddings for the best performing NDT model on each dataset. Distinct and compact clusters indicate effective class separation in the embedding space.}
    \label{fig:tsne_embeddings}
\end{figure}

Figure \ref{fig:tsne_embeddings} represents the t-SNE embeddings of the best performing NDT model for each dataset. Each subfigure corresponds to a different dataset and model configuration. 
These visualizations illustrate how the learned representations are projected into a lower dimensional space, where well separated clusters indicate better class discrimination. 
Overall, the embeddings demonstrate that the NDT model produces compact and distinct class clusters across all datasets, further confirming its ability to capture discriminative features.

\subsection{Computational Efficiency Analysis}

\subsubsection{Parameter Efficiency}
Table \ref{tab:param_analysis} presents the parameter analysis across the IMDB, SST-2, YELP, and Twitter Financial News datasets. The values indicate the total number of trainable parameters (in millions) for each model configuration, with the final column reporting the additional parameters relative to the Vanilla Transformer baseline. This absolute measure provides a clear view of the incremental parameter cost introduced by each variant. As observed, the NDT models add only a small number of extra parameters compared to DT, demonstrating that the model’s efficiency is largely maintained even as the number of components increases.

\begin{table}[h!]
\centering
\caption{Parameter Overhead Analysis}
\label{tab:param_analysis}
\begin{tabularx}{0.49\textwidth}{lccccc}
\hline
\makecell{Model\\($N_{comp}$)} & 
\makecell{IMDB\\Params\\(M)} & 
\makecell{SST-2\\Params\\(M)} & 
\makecell{YELP\\Params\\(M)} & 
\makecell{Twitter Financial\\News Params\\(M)} & 
\makecell{$\Delta$\\Params\\(M)} \\
\hline
Vanilla (--)  & 3.63 & 1.40 & 8.38 & 0.67 & -- \\
DT (2)        & 3.79 & 1.56 & 8.54 & 0.82 & $\sim0.16$ \\
NDT (2)       & 3.79 & 1.56 & 8.54 & 0.82 & $\sim0.16$ \\
NDT (3)       & 3.83 & 1.60 & 8.58 & 0.85 & $\sim0.20$ \\
NDT (4)       & 3.86 & 1.63 & 8.61 & 0.88 & $\sim0.23$ \\
\hline
\end{tabularx}
\end{table}

\subsubsection{Inference Time}
Table~\ref{tab:inf_analysis} presents the inference time comparison across the IMDB, SST-2, YELP, and Twitter Financial News datasets.
The reported values correspond to the average time (in milliseconds) required for processing a single batch of test set for each dataset. While the Differential Transformer (DT) and Non-Differential Transformer (NDT) variants incur additional latency compared to the Vanilla Transformer, the increase remains moderate relative to the model size and number of components.
As expected, inference time grows with the number of components in NDT, reflecting the added computational complexity, yet the overhead remains within a practical range.

\begin{table}[h!]
\centering
\caption{Inference Time Comparison per Batch}
\label{tab:inf_analysis}
\begin{tabular}{lcccc}
\hline
\makecell{Model\\($N_{comp}$)} & 
\makecell{IMDB\\Time (ms)} & 
\makecell{SST-2\\Time (ms)} & 
\makecell{YELP\\Time (ms)} & 
\makecell{Twitter Financial\\News Time (ms)} \\
\hline
Vanilla (--)  & 25.58 & $\sim$6.25 & 75.42  & $\sim$0 \\
DT (2)        & 38.36 & $\sim$6.25 & 121.65 & $\sim$0 \\
NDT (2)       & 40.92 & $\sim$6.25 & 126.52 & $\sim$0 \\
NDT (3)       & 48.59 & $\sim$6.25 & 158.15 & $\sim$0 \\
NDT (4)       & 58.82 & $\sim$6.25 & 197.08 & $\sim$0 \\
\hline
\end{tabular}
\end{table}

Overall, the systematic investigation of four constraint configurations reveals that purely positive, bounded combinations ([0,1]) achieve the strongest performance on binary sentiment tasks, directly supporting our hypothesis that constructive specialization outperforms destructive interference. Analysis of learned lambda values shows consistent hierarchical importance ordering across datasets, with values increasing from Layer~1 to Layer~2. The computational efficiency analysis confirms practical deployment feasibility, with NDT adding only 0.16~-~0.23M parameters relative to the Vanilla Transformer baseline while maintaining competitive inference times.

Our key contributions extend beyond empirical gains: we introduce \textit{concept-multiplexing} as an alternative theoretical foundation for multi component attention, demonstrate that purely additive combinations with positive weights can outperform subtractive approaches, and provide insights for designing attention architectures for sentiment analysis tasks. These findings challenge prevailing design principles in natural language processing and suggest that existing subtractive approaches may benefit from reconsideration through the ConPlex viewpoint. The component specialization patterns indicate that sentiment understanding emerges from simultaneous processing of multiple conceptual channels. 
The results across datasets reveal that while NDT demonstrates consistent improvements, benefits are most pronounced on binary classification tasks (IMDB, SST-2) with diminishing returns on fine grained multi class scenarios (YELP's 5 class). This pattern suggests that architectural innovations in attention mechanisms have greater impact when class boundaries are well defined.
The consistent component specialization patterns observed suggest the ConPlex framework may extend effectively to larger models.

\section{Conclusion} \label{section:conclusion}

This work introduces the \textbf{Non-Differential Transformer (NDT)}, a novel attention mechanism that challenges the noise cancellation hypothesis underlying subtractive attention approaches. Through our \textbf{concept-multiplexing (ConPlex)} framework, we demonstrate that constructive integration of specialized attention components achieves superior performance compared to subtractive mechanisms in sentiment analysis tasks. Our experimental evaluation across four diverse datasets, viz. \textit{IMDB Movie Reviews}, \textit{SST-2}, \textit{YELP Reviews}, and \textit{Twitter Financial News}, provides compelling evidence for additive attention combinations, with NDT consistently outperforming both Vanilla Transformers and the Differential Transformer by up to 1.36\% on IMDB, 0.32\% on SST-2, 0.32\% on YELP Reviews, and 1.00\% on Twitter Financial News.
Future work should extend NDT to deeper architectures and larger scale pretraining to validate scalability.

\bibliographystyle{IEEEtran}
\bibliography{references}

\end{document}